\documentclass[aps,pra,twocolumn,reprint,superscriptaddress]{revtex4-2}

\usepackage{hyperref}
\usepackage{graphicx}  % needed for figures
\usepackage{bm}        % for math
\usepackage{amssymb}   % for math
\usepackage{xspace}
\usepackage{verbatim}
\usepackage{color}
\usepackage{soul}
\usepackage{subfigure}
\usepackage[export]{adjustbox}
\usepackage{dcolumn}% Align table columns on decimal point
\usepackage{amsthm,stmaryrd,mathtools}
\usepackage{txfonts}
\usepackage{braket}
\usepackage{amsmath}
\usepackage{xcolor}
\usepackage{array}
\usepackage{multirow}
\usepackage{tabularx}
\usepackage{booktabs}
\usepackage{lipsum}

\begin{document}

\title{Nonlinearity-enhanced quantum sensing in Stark probes}

\author{Rozhin Yousefjani}%
\thanks{These authors have contributed equally to this paper.}
\affiliation{Institute of Fundamental and Frontier Sciences, University of Electronic Science and Technology of China, Chengdu 611731, China}

\author{Xingjian He}%
\thanks{These authors have contributed equally to this paper.}
\affiliation{Institute of Fundamental and Frontier Sciences, University of Electronic Science and Technology of China, Chengdu 611731, China}

\author{Angelo Carollo}%
\affiliation{Dipartimento di Fisica e Chimica “E. Segr\`{e}”, Group of Interdisciplinary Theoretical Physics, Universit`a degli Studi di Palermo, I-90128 Palermo, Italy}

\author{Abolfazl Bayat}%
\email{abolfazl.bayat@uestc.edu.cn}
\affiliation{Institute of Fundamental and Frontier Sciences, University of Electronic Science and Technology of China, Chengdu 611731, China}
\affiliation{Key Laboratory of Quantum Physics and Photonic Quantum Information, Ministry of Education, University of Electronic Science and Technology of China, Chengdu 611731, China}

\begin{abstract}
Stark systems in which a linear gradient field is applied across a many-body system have recently been proposed for quantum sensing. 
Here, we explore sensing capacity of Stark probes, in both single-particle and many-body interacting systems, for estimating nonlinear forms of the gradient fields. 
Our analysis reveals that, this estimation can achieve super-Heisenberg scaling precision that grows linearly by increasing the nonlinearity.  
Specifically, we find a universal algebraic relation between the scaling of the precision and the degree of the nonlinearity.
This universal behavior remains valid in both single-particle and many-body interacting probes and reflects itself in the properties of the phase transition from an extended to a localized phase, obtained through establishing a comprehensive finite-size scaling analysis.
Considering a parabolic gradient potential composed of both linear and nonlinear fields, we used multi-parameter estimation methodology to estimate the components of the gradient potential. 
The phase diagram of the system is determined in terms of both linear and nonlinear gradient fields showing how the nonlocalized phase turns into a localized one as the Stark fields increase. The sensing precision of both linear and nonlinear Stark fields follows the same universal algebraic relation that was found for the case of single parameter sensing. We demonstrate that simple and experimentally available measurements can reach the theoretical precision bounds.  
Finally, we show that quantum enhanced sensitivity is still achievable even when we incorporate the preparation time of the probe into our resource analysis.    
\end{abstract}

\maketitle
%%%%%%%%%%%%%%%%%%%%%%%%%%%%%%%%%%%%%%%%%%%%%%%%%%%%%%%%%%%%%%%%%%%%%%%%%%%%%%%%%%%%%%%%%%%%%%%%%%%%%%%%%%%%%%%%%%%%%%%%%%%%%%%%%%%%%%%%%%%%%%%%%%%%%%%%%%%%%%%%%%%%%%%%

\section{Introduction}
Due to extreme sensitivity to variations in environment, quantum systems can serve as sensors whose precision can exceed their classical counterparts~\cite{roy2008exponentially,paris2009quantum,banaszek2009quantum,braun2018quantum,boixo2007generalized,beau2017nonlinear,degen2017quantum,YOUSEFJANI201780}.
This superiority manifests itself in the uncertainty of their estimation, 
quantified by variance, which scales as ${~}L^{-\beta}$, where $L$ is probe size and $\beta$ is scaling exponent~\cite{rao1992information,braunstein1994statistical,cramer1999mathematical}.
The best performance of the classical probes is limited to the standard quantum limit, namely $\beta{=}1$, determined by the central limit theorem.
More favorable estimation precision with $\beta{>}1$ might be achievable through harnessing quantum features, e.g. entanglement, which is known as quantum-enhanced sensitivity. 
Originally, such enhancement was discovered for a probe made of non-interacting particles initialized in a certain type of entangled states, known as Greenberger-Horne-Zeilinger (GHZ) states~\cite{greenberger1989going,giovannetti2004quantum,giovannetti2006quantum,giovannetti2011advances,frowis2011stable,demkowicz2012elusive,wang2018entanglement,kwon2019nonclassicality}. 
In GHZ-based sensors, the scaling of uncertainty improves to the Heisenberg limit, i.e. $\beta{=}2$. 
Nevertheless, susceptibility of those probes to decoherence~\cite{albarelli2018restoring,dur2004stability,demkowicz2012elusive,matsuzaki2011magnetic,shaji2007qubit,bhattacharyya2024tunable} and inter-particle interaction~\cite{de2013quantum,pang2014quantum} puts serious challenges for scaling up.
Moreover, the precision of GHZ-based sensors with non-interacting particles is strictly bounded by the Heisenberg limit.
To overcome these challenges, strongly correlated many-body systems, have been proposed as alternative types of sensors in which interaction between particles plays a central role.
In particular, quantum criticality has been identified as a resource for achieving quantum-enhanced sensitivity. 
Several types of criticality have been used for achieving quantum enhanced sensitivity, including first-order~\cite{raghunandan2018high,heugel2019quantum,yang2019engineering}, second-order~\cite{zanardi2006ground,zanardi2007mixed,gu2008fidelity,zanardi2008quantum,invernizzi2008optimal,gu2010fidelity,gammelmark2011phase,skotiniotis2015quantum,rams2018limits,wei2019fidelity,chu2021dynamic,liu2021experimental,montenegro2021global,mirkhalaf2021criticality,di2021critical,Salvia2023Critical}, 
dissipative~\cite{fernandez2017quantum,baumann2010dicke,baden2014realization,klinder2015dynamical,rodriguez2017probing,fitzpatrick2017observation,fink2017observation,ilias2022criticality,Ilias2023Criticality,Alipour2014Quantum}, 
time crystals~\cite{yousefjani2024DTC,montenegro2023quantum,iemini2023floquet}, Floquet~\cite{mishra2021driving,mishra2022integrable}, topological~\cite{budich2020non,sarkar2022free,koch2022quantum,yu2022experimental} and Stark phase transitions~\cite{he2023stark,Yousefjani2023}. 
The notion has also been generalized to non-Hermitian open quantum systems~\cite{wiersig2014enhancing,Budich2020Non-Hermitian,mcdonald2020exponentially,sarkar2023quantumenhanced}.

Unlike GHZ-based sensors, the precision of criticality-based many-body probes is not bounded and super-Heisenberg scaling, namely $\beta{>}2$, may also be achievable~\cite{boixo2007generalized,gu2008fidelity,roy2008exponentially,beau2017nonlinear,rams2018limits,wei2019fidelity,rubio2021global}.
However, the stringent requirement of initializing these probes in their ground state near the critical point, via, for instance, the adiabatic evolution, imposes difficulties in accessing this enhancement.
As such, the possibility of exploiting the criticality turns to a hot debate in both theoretical~\cite{rams2018limits} and experimental~\cite{liu2021experimental,yu2022experimental,ilias2024criticality} arena. Recently, Stark many-body probes have been introduced for measuring linear gradient fields with an unprecedented precision of $\beta{\simeq}6$~\cite{he2023stark,Yousefjani2023}. 
Emerging onsite off-resonance energy in the presence of a gradient field localizes the wave function of the particles even in the presence of strong interaction. 
This interesting phenomenon, known as Stark localization~\cite{wannier1960wave}, has been subject of recent studies~\cite{fukuyama1973tightly,holthaus1995random,kolovsky2003bloch,kolovsky2008interplay,kolovsky2013wannier,van2019bloch,van2019bloch,schulz2019stark,wu2019bath,bhakuni2020drive,bhakuni2020stability,yao2020many,chanda2020coexistence,taylor2020experimental,wang2021stark,zhang2021mobility,guo2021stark,yao2021many,doggen2022many,zisling2022transport,burin2022exact,bertoni2024local,lukin2022many,vernek2022robustness,Doggen2021Stark,Sahoo2024Localization} and has been observed in different experimental platforms including ion traps~\cite{morong2021observation}, optical lattices~\cite{preiss2015strongly,kohlert2021experimental}, and superconducting simulators~\cite{karamlou2022quantum}.
There are two key features that make the Stark probes very distinct from the other many-body sensors. First, their best performance is obtained for small fields which most probes fail to estimate. 
Second, the Stark transition takes place across the whole spectrum and thus the requirement of precise preparation of the ground state is relaxed. While the localization properties of nonlinear Stark systems have been studied in several works~\cite{yao2021many,taylor2020experimental,wang2021stark,bertoni2024local}, their potential as quantum sensors have not yet been explored.  

In this work, we systematically address this issue by investigating the sensing capability of the Stark probe for estimating different forms of gradient fields. 
We study this problem in single-particle and many-body interacting probes. 
In the following, we sketch our main findings. 
For a general nonlinear gradient field,
our comprehensive analysis reveals a simple universal algebraic relation between the degree of nonlinearity in the gradient field and the scaling exponent of the QFI, namely $\beta$. 
Interestingly, the scaling exponent $\beta$ and all critical parameters of the system grow linearly with nonlinearity exponent. 
We also investigate Stark systems under the impact of both linear and quadratic fields, where the problem naturally lies in the context of multi-parameter sensing~\cite{Ragy2016Compatibility,di2021critical,Liu2020Quantum,Yousefjani2017Estimating,Carollo2019}.
The phase diagram of the system which denotes the transition from extended to localized phase is specified by all elements of the quantum Fisher information matrix.
Quantum-enhanced sensitivity is achievable throughout the extended phase. 
After fully characterizing the critical properties of the probe through finite-size scaling analysis, we propose proper sets of projective measurements that allow us to capture the obtained quantum-enhanced sensitivity. 
Analyzing the available resources in a typical sensing setup reveals that the offered enhanced sensitivity by the Stark probes is always achievable even after considering the initialization cost.
\\
The paper is organized as follows. We begin by recapitulating the theory of estimation in both single- and multi-parameter levels and laying the analytic arguments of relevance to this study in section~\ref{S.PET}.
In section~\ref{S.NSP}, we present the results for single-parameter estimation using single-particle probe~\ref{SubS.NSP-SPP} and many-body interacting probe~\ref{SubS.NSP-MBP}. 
The results for estimating a parabolic gradient potential, and hence multi-parameter estimation, are reported in section~\ref{S.NSP-MPE}. 
In~\ref{SubS.NMP-SPP}, the sensing power of a single-particle probe for estimating both linear and nonlinear fields is investigated. 
\ref{SubS.NMP-MBP} deals with the multi-parameter estimation using many-body interacting probes. 
Additionally, the study is strengthened by including analysis for optimal measurement~\ref{SubS.MPE-OM}, resource analysis~\ref{SubS.MPE-RA}, and simultaneous estimation~\ref{SubS.MPE-SE} for both single-particle and many-body interacting probes.  
Finally, the paper has been summarized in section~\ref{ConClusion}.

\section{Parameter estimation theory}\label{S.PET}
In this section, we briefly review the concepts of estimation theory for single-parameter and multi-parameter quantum sensing~\cite{Ragy2016Compatibility,di2021critical,Liu2020Quantum,Yousefjani2017Estimating}. 
\\
\subsection{Single-parameter estimation}\label{SubS.SPE}
In single-parameter sensing one deals with a quantum probe whose density matrix $\rho(h)$ varies with an under scrutiny parameter $h$. 
Sensing this parameter demands performing an appropriate measurement on the probe. 
Implementing a set of positive operator-valued measurement (POVM) $\{\Pi_{k}\}$ results in probability distribution $p_k(h){=}\mathrm{Tr}[\Pi_{k}\rho(h)]$ as the outcomes of the measurement and, hence, Classical Fisher Information (CFI) $\mathcal{F}_{C}(h){=}\sum_{k}p_{k}(h)(\partial_{h}\ln p_{k}(h))^2$.
The CFI determines the lower bound on the 
precision of estimation, quantified by variance $\delta h^2{=}\langle h^2 \rangle {-} \langle h \rangle^2$, through Cram\'{e}r-Rao inequality as $\delta h^2{\geqslant}1{/}M\mathcal{F}_{C}(h)$. Here $M$ is the number of samples.
The ultimate precision limit achievable by the quantum probe is determined using Quantum Fisher Information (QFI), obtained by optimizing the CFI over all possible POVMs, namely $\mathcal{F}_Q(h){=}\max_{\{\Pi_{k}\}}\mathcal{F}_C(h)$.
This results in a tighter bound for variance known as the quantum Cram\'{e}r-Rao bound 
\begin{equation}
    \delta h^2 \geqslant \frac{1}{M \mathcal{F}_{Q}(h)}.
\end{equation}
This inequality is always saturable in the
limit of a large number of samples $M$ and using an optimal measurement basis given by the eigenvectors of the 
Symmetric Logarithmic Derivative (SLD) operator denoted as $L(h)$. Note that SLD as a Hermitian operator satisfies 
\begin{equation}\label{Eq.SLD-Single}
\partial_{h}\rho(h)=\frac{\rho(h)L(h)+L(h) \rho(h)}{2},
\end{equation}
with $\partial_{h}{=}\partial/\partial h$ and leads to a closed form for the QFI as
\begin{equation}\label{Eq.QFI-single}
 \mathcal{F}_{Q}(h) = {\rm Tr}[\rho(h) L(h)^2].  
\end{equation}
For pure states $\rho_{h}{=}|\psi(h)\rangle \langle\psi(h)|$, Eq.(\ref{Eq.QFI-single}) is simplified to
\begin{equation}\label{Eq.QFI-Single-Pure}
{\mathcal{F}_{Q}}({h})=4\left[\langle \partial_{h}\psi|\partial_{h}\psi\rangle-\langle \partial_{h}\psi|\psi\rangle \langle \psi|\partial_{h}\psi\rangle \right].
\end{equation} 
\\
\subsection{Multi-parameter estimation}\label{SubS.MPE}
In multi-parameter sensing, a set of $p{\geqslant}2$ unknown parameters as $\boldsymbol{h}=\{h_1,h_2,...,h_p\}$ are under scrutiny.
For $\rho(\boldsymbol{h})$, the precision of estimating $\boldsymbol{h}$ is quantified through a $p\times p$ covariance matrix $\boldsymbol{Cov}(\boldsymbol{h})$ with elements as $(\boldsymbol{Cov})_{ij}(\boldsymbol{h}){=} \langle  h_i h_j \rangle - \langle  h_i \rangle \langle h_j \rangle$.
For a set of POVM $\{\Pi_k\}$ and outcomes $p_k(\boldsymbol{h}){=}{\rm Tr}[\Pi_k\rho(\boldsymbol{h})]$,
the covariance matrix satisfies the multi-parameter  Cram\'{e}r-Rao inequality as
\begin{equation}\label{Eq.CCRBM}
\boldsymbol{Cov}(\boldsymbol{h})\geqslant \frac{1}{M}\boldsymbol{\mathcal{F}_{C}}^{-1}(\boldsymbol{h}),
\end{equation}
where $\boldsymbol{\mathcal{F}_{C}}$ is $p\times p$ CFI matrix with elements as
\begin{equation}\label{Eq.CCRBM_elements}
(\boldsymbol{\mathcal{F}_{C}})_{ij}(\boldsymbol{h}) = \sum_k
p_k(\boldsymbol{h})\Big(\partial_i \ln p_k(\boldsymbol{h})\Big)\Big(\partial_j \ln p_k(\boldsymbol{h})\Big).
\end{equation}
Here $\partial_i{=}\partial/\partial h_i$. The ultimate precision limit in this case is   given by quantum multi-parameter Cram\'{e}r-Rao inequality as
\begin{equation}\label{Eq.QCRBM}
\boldsymbol{Cov}(\boldsymbol{h})\geqslant \frac{1}{M}\boldsymbol{\mathcal{F}_{Q}}^{-1}(\boldsymbol{h}),
\end{equation}
where $\boldsymbol{\mathcal{F}_{Q}}$ is $p\times p$ QFI matrix. 
The elements of the QFI matrix using the corresponding SLD operators, denoted as $L_i$ concerning $h_i$, are given by~\cite{Helstrom1976}
\begin{equation}\label{Eq.QFIM_SLD}
(\boldsymbol{\mathcal{F}_{Q}})_{ij}(\boldsymbol{h}) = \frac{1}{2} \text{Tr}[\rho({\boldsymbol{h}})(L_{i}L_{j} + L_{j}L_{i})].
\end{equation}
For pure states, namely $\rho({\boldsymbol{h}}){=}|\psi(\boldsymbol{h})\rangle \langle\psi(\boldsymbol{h})|$, the QFI matrix elements are simplified to
\begin{equation}\label{Eq.QFIM}
(\boldsymbol{\mathcal{F}_{Q}})_{ij}(\boldsymbol{h})=4\rm{Re}\left[\langle \partial_{\textit{i}}\psi|\partial_{\textit{j}}\psi\rangle-\langle \partial_{\textit{i}}\psi|\psi\rangle \langle \psi|\partial_{\textit{j}}\psi\rangle \right].
\end{equation}
In the case of multi-parameter estimation, the bound Eq.~(\ref{Eq.QCRBM}) is not tight. Intuitively, this is due to the non-commutativity of the optimal measurements associated to different parameters. When $[L_i,L_j]{=}0$, for all choices of $i$ and $j$, the SLD operators are simultaneously diagonalizable, and the saturation condition is satisfied. 
It turns out that weaker condition ${\rm Tr}(\rho({\boldsymbol{h}})[L_{i},L_{j}]){=}0$ for all $i$ and $j$ is the necessary and sufficient condition for saturation of the multi-parameter quantum Cram\'er-Rao bound~\cite{Ragy2016Compatibility}. 
\\

Note that Eq.~(\ref{Eq.QCRBM}) is a matrix inequality and to obtain a scalar inequality, one can multiply both sides with a positive weight matrix $\boldsymbol{W}$ and compute the trace 
\begin{equation}\label{Eq.WeightedQRB}
\text{Tr}[\boldsymbol{W}\boldsymbol{Cov}(\boldsymbol{h})] \geqslant M^{-1} \text{Tr}[\boldsymbol{W}\boldsymbol{\mathcal{F}_{Q}}^{-1}(\boldsymbol{h})]. 
\end{equation}
The weight matrix $\boldsymbol{W}$ can be chosen to add any combination of the elements of the covariance matrix as a measure of total uncertainty on the left side of the above inequality. 
In particular, one can choose $\boldsymbol{W}{=}\mathbb{I}$ for which the total uncertainty becomes the summation of variances of all the parameters and is bounded through $\sum^p_{i=1}\delta h_i^2\geqslant M^{-1}{\rm Tr}[\boldsymbol{\mathcal{F}_{Q}}^{-1}(\boldsymbol{h})]$. 
If one is only interested in the precision of estimating $h_{i}$, the weight matrix $\boldsymbol{W}$ needs to be chosen with only one non-zero element, namely $(\boldsymbol{W})_{ii}{=}1$. 
In this case, Eq.~(\ref{Eq.WeightedQRB}) reduces to $\delta h_{i}^2{\geqslant} M^{-1}(\boldsymbol{\mathcal{F}_{Q}}^{-1}(\boldsymbol{h}))_{ii}$.
Note that this is different from standard single-parameter sensing in which only the parameter under scrutiny is unknown.  
In general, the non-zero correlations between the unknown parameters in $\boldsymbol{h}$, quantified through the off-diagonal elements of the QFI matrix, affect the sensing precision of each $h_i$. 
However, the simultaneous estimation performance always overcomes one of the separable estimations by taking advantage of the available resources in sensing tasks, namely the number of sampling $M$.
While in a simultaneous scenario, the total uncertainty $\sum_{i=1}^{p}\delta h_i^2$ is lower bounded by  $M^{-1}{\rm Tr}[\boldsymbol{\mathcal{F}_{Q}}^{-1}(\boldsymbol{h})]$, in the separable scenario this quantity reduces to $(M/p)^{-1}{\rm Tr}[\boldsymbol{\mathcal{F}_{Q}}^{-1}(\boldsymbol{h})]$ as one needs to implement the whole estimation protocol $p$ times to sense all the parameters.     
%%%%%%%%%%%%%%%%%%%%%%%%%%%%%%%%%%%%%%%%
\section{Nonlinear Gradient Field Sensing:~\\~Single-Parameter Estimation}\label{S.NSP}
Recently, the Stark probe has been exploited for sensing the linear gradient magnetic field in both single-particle and many-body levels~\cite{he2023stark,Yousefjani2023}. 
Three main results have been observed. 
First,  all the eigenstates of the system show quantum-enhanced sensitivity, in terms of the system size $L$,  with super-Heisenberg precision in the extended regime and transition point. 
Second, in the localized regime, the sensitivity becomes size independent and the system shows universal behavior. 
Third, the phase transition from the extended to the localized phase is described by a continuous phase transition formalism which implies the emergence of a diverging length scale at the transition point. 
Here we aim to reveal the effect of nonlinearity in the gradient potential on the sensing power of both single-particle and many-body interacting probes.  

\subsection{Single-particle probe}\label{SubS.NSP-SPP}
We begin by recapitulating the physics of Stark localization transition at single-particle levels. 
Let's consider a one-dimensional probe with $L$ sites in which a single particle can tunnel between neighboring sites with a rate $J{>}0$, in the presence of 
gradient potential $V_i{=}hi^\gamma$.
The total Hamiltonian of the system reads
\begin{equation}\label{Eq:Stark_Hamiltonian}
H = J\sum_{i=1}^{L-1} \left( \vert i\rangle\langle i+1\vert + h.c.  \right) + \sum_{i=1}^{L}  V_{i} \vert i\rangle\langle i\vert.
\end{equation}
In the absence of gradient potential, i.e. $V_{i}{=}0$, the Hamiltonian describes an integrable model and can be easily diagonalized to extended Bloch eigensystem as
\begin{eqnarray}\label{Eq:eigensystem}
	E_{k}&=&-2J{\rm cos}(\frac{k\pi}{L+1})\cr
	|E_k\rangle&=&\sqrt{\frac{2}{L+1}}\sum_{j=1}^L (-1)^j\sin\left(\frac{jk\pi}{L+1}\right)\vert j\rangle,
\end{eqnarray}
where index $k{=}1,{\cdots},L$ counts all the eigenstates of the system. 
Clearly, the eigenstates are extended across the entire system, known as the extended phase~\cite{holthaus1995random,kolovsky2008interplay}.
In the presence of gradient fields $V_{i}{\neq}0$,  the off-resonant energy splitting at each site tends to localize the wave function of the particle, known as Stark localization.
It is well known that, the transition from the extended to the localized phase takes place across the entire spectrum for a gradient field that approaches zero ($h_c{\rightarrow}0$) in the thermodynamic limit ($L{\rightarrow}\infty$)~\cite{kolovsky2008interplay}. 
\\
\begin{figure}[t]
\includegraphics[width=\linewidth]{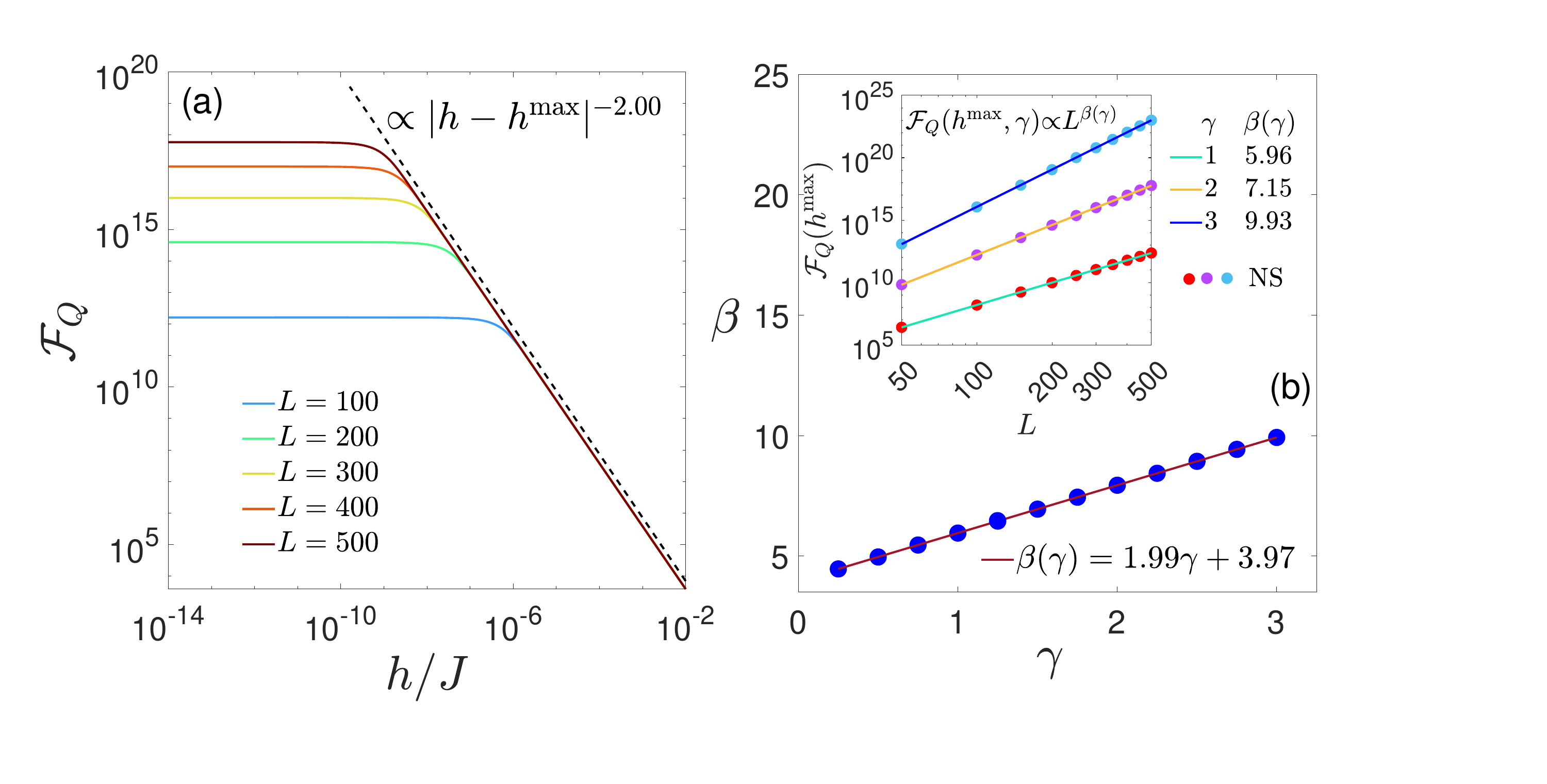}  
\includegraphics[width=\linewidth]{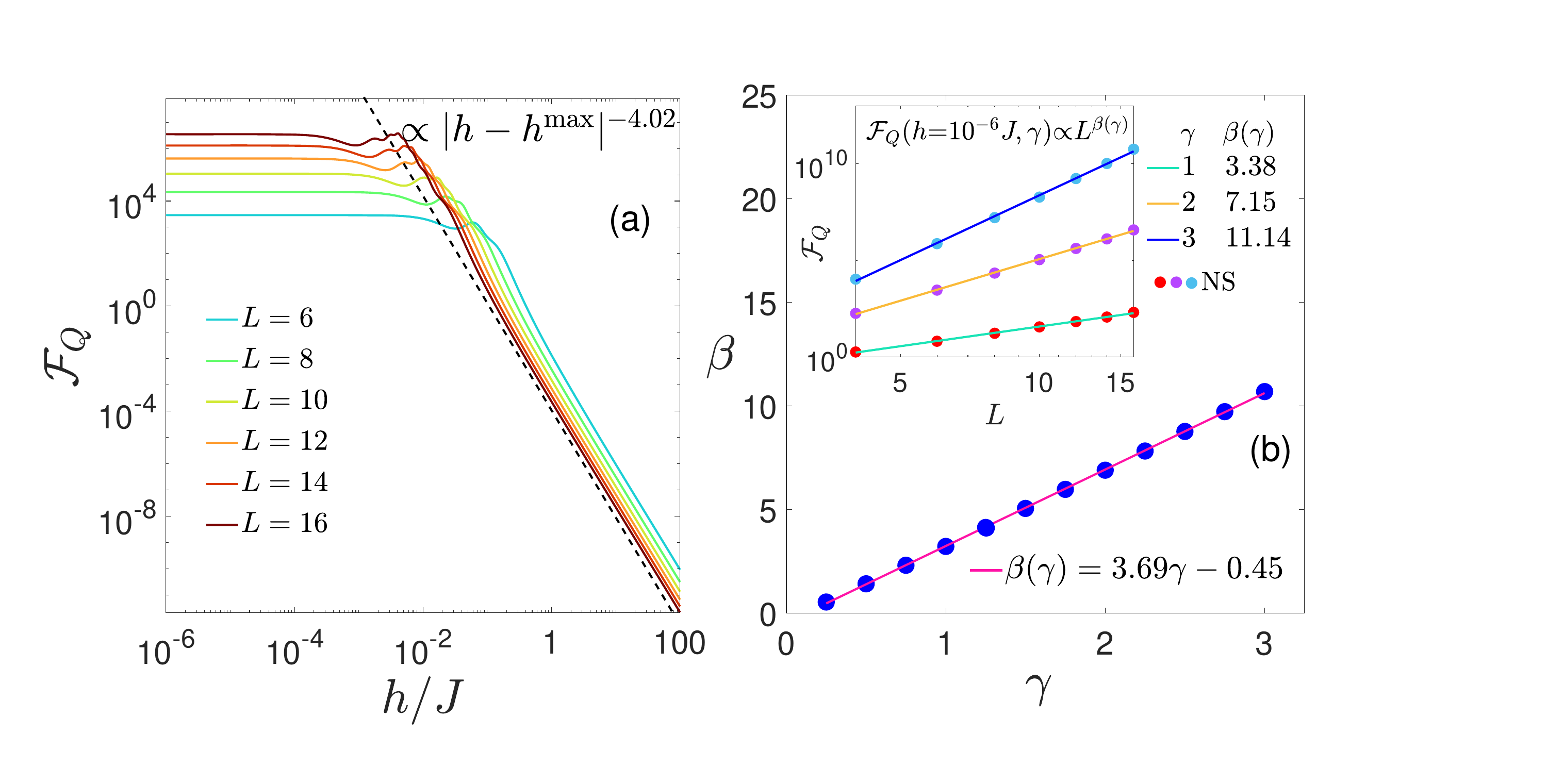} 
\caption{(a) QFI, $\mathcal{F}_Q$ as a function of the gradient magnitude $h$ in a single-particle probe with nonlinear gradient potential $V_i{=}hi^2$. The probes of different sizes  are prepared in the corresponding ground state of Eq.~(\ref{Eq:Stark_Hamiltonian}). The universal algebraic reduction of the QFI in the localized phase is described by $\mathcal{F}_{Q}{\propto}|h-h^{\max}|^{-\alpha}$ with $\alpha{=}2.00$ (dashed line).
Inset of (b) is the maximum of the QFI, $\mathcal{F}_{Q}(h^{\max},\gamma)$ as a function of $L$ for various $\gamma$'s. The numerical simulation (NS) are properly described by the fitting function $\mathcal{F}_{Q}(h^{\max},\gamma){\propto}L^{\beta(\gamma)}$ with corresponding reported exponent $\beta(\gamma)$. 
(b) The extracted exponent $\beta$'s as a function of $\gamma$ for single-particle probe. The NS (markers) are fitted by the linear function $\beta(\gamma){=}a\gamma{+}b$ (solid line), with $(a,b){\simeq}(1.99,3.97)$.
(c) Finite-size scaling analysis for the curves in panel (a). The critical parameters are tuned based on the reported $(h_c,\alpha,\nu)$. (d) the critical exponent $1{/}\nu$ as well as $\alpha{/}\nu$ versus $\gamma$. The NS are well represented by algebraic function as ${\simeq}a\gamma{+}b$ with reported $(a,b)$. }\label{fig:nonLinearEffect_SP}
\end{figure}

To assess the performance of the probe, in Fig.~\ref{fig:nonLinearEffect_SP} (a) we plot the QFI as a function of the gradient field in systems of different sizes $L{\in}\{101,\cdots,501\}$ that are initialized in the ground states of Eq.~(\ref{Eq:Stark_Hamiltonian}) with $V_i{=}hi^2$. 
Obviously, the QFI behaves differently in both extended and localized phases. 
One observes a size-dependent plateau in the extended phase that survives over an interval of $h$. The maximum values of the QFI happen at $h^{\max}$ that tends to smaller values by enlarging the chain and vanish in the thermodynamic limit, namely $h^{\max}{\rightarrow}h_c{=}0$. 
In the localized phase, the finite-size effect disappears, and a universal algebraic decrease of the QFI as $\mathcal{F}_{Q}{\propto}|h-h^{\max}|^{-\alpha}$ with $\alpha{=}2.00$ is evidenced. 
Note that, our analysis shows that the universal decay of the QFI in the localized phase is general and independent of $\gamma$. In particular for other degrees of nonlinearity $\gamma{\in}[0.25,\cdots,3]$, we obtain $\alpha{=}2.00$.
The fact that the transition between extended and localized phases in the ground state of Eq.~(\ref{Eq:Stark_Hamiltonian}) happens in $h^{\max}{\rightarrow}0$ allows us to describe the physics of the system using methods for a slightly perturbed integrable model.
Relying on this, we establish an analytical analysis for extracting the scaling behavior of $\mathcal{F}_{Q}$.
Expressing Eq.~(\ref{Eq:Stark_Hamiltonian}) as $H{=}H_{0}{+}h H_1$, with $H_0$ as the first term of the Hamiltonian and $H_1{=}\sum_{i=1}^{L}i^\gamma |i\rangle\langle i|$, allows to take advantage of the perturbation theory and hence rewrite the QFI in Eq.~(\ref{Eq.QFI-Single-Pure}) in terms of the spectrum Eq.~(\ref{Eq:eigensystem})  as    
\begin{equation}\label{Eq.QFI_analytical1}
    \mathcal{F}_{Q}(h^{\max}{\rightarrow} 0,\gamma)=4\sum_{k\neq 1}\frac{|\langle E_{k}|H_1|E_1\rangle|^2}{|E_{k}-E_{1}|^2}.
\end{equation}
Straightforward simplifications result in     
\begin{equation}\label{Eq.QFI_analytical2}
    \mathcal{F}_{Q}(h^{\max}{\rightarrow} 0,\gamma)
    =\frac{4}{J^2(L+1)^2}\sum_{k\neq 1}\frac{\mathcal{N}(k)}{\mathcal{D}(k)}>\frac{4}{J^2(L+1)^2}\frac{\mathcal{N}(k{=}2)}{\mathcal{D}(k{=}2)},
\end{equation}
in which $\mathcal{N}(k){=}[\sum_{i=1}^L i^{\gamma}{\rm sin}(\frac{ik\pi}{L+1}){\rm sin}(\frac{i\pi}{L+1})]^2$ and $\mathcal{D}(k){=}[{\rm cos}(\frac{k\pi}{L+1}){-}{\rm cos}(\frac{\pi}{L+1})]^2$. In writing the nonequality, we use the fact that $\mathcal{N}(k){/}\mathcal{D}(k){>}0$ and $\mathcal{N}(k{=}2){/}\mathcal{D}(k{=}2)$ has the main contribution in $\sum_{k{\neq 1}}$.
A closed form for the most right term in Eq.~(\ref{Eq.QFI_analytical2}) can be obtained using $\mathcal{N}(K{=}2){\simeq}(1{-}L^{-0.01})L^{2\gamma{+}2}{/(\gamma{+}1)^2}$ and $\mathcal{D}(K{=}2){\simeq}\frac{9\pi^4}{4(L{+}1)^4}$.
Regarding the scaling of the QFI with respect to $L$, one can simply obtain 
\begin{equation}\label{Eq.AnalyticalQFI}
    \mathcal{F}_{Q}(h^{\max}{\rightarrow} 0,\gamma) \propto L^{2\gamma+3.99}(L^{0.01} - 1).
\end{equation}
To provide numerical support for the above analysis one can plot the maximum values of the QFI, $\mathcal{F}_{Q}(h^{\max},\gamma)$,  versus system size $L$ for various $\gamma$'s.
As it has been shown in the inset of  Fig.~\ref{fig:nonLinearEffect_SP} (b),
the numerical simulation (NS) are properly describe by a fitting function of the form $\mathcal{F}_{Q}{\propto}L^{\beta(\gamma)}$. 
In Fig.~\ref{fig:nonLinearEffect_SP} (b), we report the extracted $\beta$'s as a function of $\gamma$. 
The results are obtained from analyzing systems of sizes $L{\in}\{100,\cdots,500\}$ and are properly described by the fitting function
\begin{equation}\label{Eq.Beta}
 \beta = a\gamma {+}b,   
\end{equation}
with $(a,b){\sim}(1.99,3.97)$. 
This yields in 
\begin{equation}
    \mathcal{F}_{Q}(h^{\max}{\rightarrow} 0,\gamma) \propto L^{1.99\gamma{+}3.97},
\end{equation}
which shows perfect congruence with the result of the analytical analysis Eq.~(\ref{Eq.AnalyticalQFI}).  
\\
To identify the transition properties of our nonlinear Stark probe, we establish a finite-size scaling analysis, relying on this assumption that the Stark transition is of continuous phase transition type. Therefore one expects that $\mathcal{F}_{Q}$ adapt the following ansatz 
\begin{equation}
\mathcal{F}_{Q}(h,\gamma)=L^{\alpha(\gamma)/\nu(\gamma)}\mathcal{G}(L^{1/\nu(\gamma)}(h-h_c(\gamma))),
\end{equation}
with $\mathcal{G}(\bullet)$ as an arbitrary function and $\alpha(\gamma)$ and $\nu(\gamma)$ as critical exponents for the typical $\gamma$.
Following the finite-size scaling framework, if one plot $L^{-\alpha(\gamma)/\nu(\gamma)}\mathcal{F}_{Q}$  as a function of $L^{1/\nu(\gamma)}(h-h_c(\gamma))$, the curves for different sizes collapse on one graph for appropriate exponents and $h_c$.
In Fig.~\ref{fig:nonLinearEffect_SP} (c) we present the optimal data collapse for $\gamma{=}2$ that is obtained by tuning the critical parameters as $(h_c,\alpha,\nu){=}(3.30{\times}10^{-12}, 2.00,0.25)$. 
Our analysis shows that, while the exponent $\alpha$, which also determines the decay rate in the localized phase (dashed line in Fig.~\ref{fig:nonLinearEffect_SP} (a)), is universal and independent of $\gamma$, the exponent $\nu(\gamma)$ decreases by increasing nonlinearity. In Fig.~\ref{fig:nonLinearEffect_SP} (d) we report $1{/}\nu$ versus $\gamma$ that clearly satisfy the algebraic pattern $1{/}\nu{=}a\gamma{+}b$ with $(a,b){=}(1.01,1.97)$.
Moreover, $\alpha{/}\nu$ as a function of $\gamma$ has been plotted which shows $\alpha{/}\nu{\simeq}\beta$.
The relationship between QFI scaling $\beta$ and critical exponents $(\alpha,\nu)$ for the linear Stark probe has been discussed in Refs.~\cite{he2023stark,Yousefjani2023}. 
The above analysis suggest the universality of this relationship concerning $\gamma$.

\subsection{Many-body interacting probe}\label{SubS.NSP-MBP}
\begin{figure}[t]
\includegraphics[width=\linewidth]{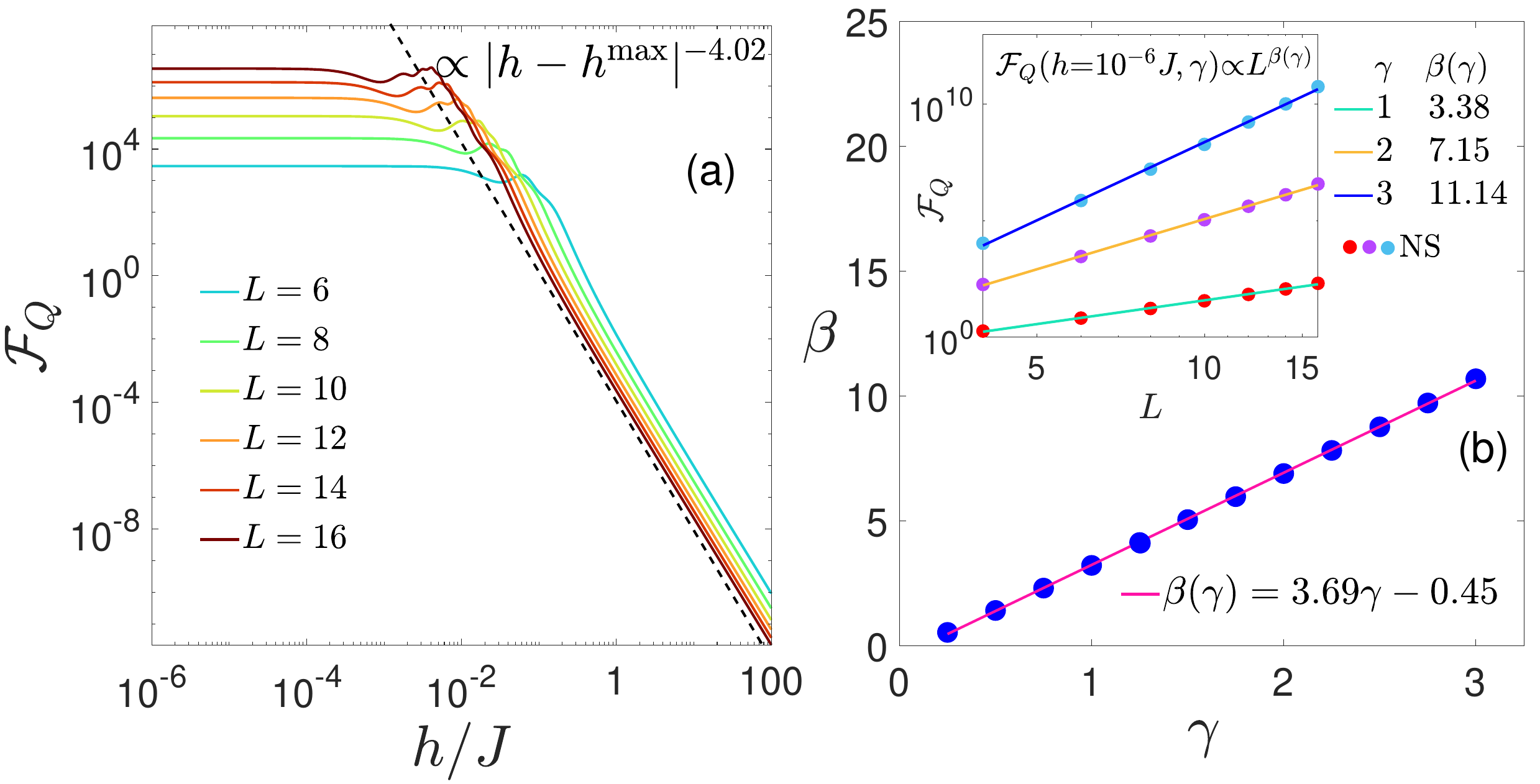} 
\caption{(a) QFI, $\mathcal{F}_Q$, as a function of the gradient magnitude $h$ in a many-body interacting probe with nonlinear gradient potential $V_i{=}hi^2$. The probes of different sizes  are prepared in the corresponding ground state of Eq.~(\ref{Eq:Stark_MB_Hamiltonian}). The universal algebraic reduction of the QFI in the localized phase is described by $\mathcal{F}_{Q}{\propto}|h-h^{\max}|^{-\alpha}$ with $\alpha{\simeq}4$ (dashed line).
Inset of (b) is the QFI in $h{=}10^{-6}J$ as a function of $L$ for various $\gamma$'s. The NS are properly described by the fitting function $\mathcal{F}_{Q}(h{=}10^{-6}J,\gamma){\propto}L^{\beta(\gamma)}$ with corresponding reported exponent $\beta(\gamma)$. 
(b) The extracted exponent $\beta$'s as a function of $\gamma$ for many-body interacting probe. The NS (markers) are fitted by the linear function $\beta(\gamma){=}a\gamma{+}b$ (solid line), with $(a,b){\sim}(3.69,-0.45)$.
}\label{fig:nonLinearEffect_MB}
\end{figure}

The Stark localization is not limited to systems of single-particle and has also been observed in many-body interacting systems~\cite{Doggen2021Stark}.
Here we consider a probe of size $L$ in the half-filling regime where $N{=}L/2$ particles interact with each other while exposed by the gradient potential $V_i{=}hi^{\gamma}$. The total Hamiltonian is
\begin{equation}\label{Eq:Stark_MB_Hamiltonian}
\begin{aligned}
H = J\sum_{i=1}^{L-1}\boldsymbol{\sigma_i\cdot\sigma_{i+1}} + \sum_{i=1}^{L} V_{i} \sigma_i^z,
\end{aligned}
\end{equation}
where $\boldsymbol{\sigma_i}{=}(\sigma_i^x,\sigma_i^y,\sigma_i^z)$ and $\sigma_i^{x,y,z}$ are the Pauli operators acting at site $i$. The half filling subspace is defined by  $\langle S_z^{tot} \rangle{=}\langle\sum_{i}^{L}\sigma^{z}_{i=1}\rangle{=}0$.
In this case, the many-body localization is a direct result of the competition between interaction and the gradient potential $V_i$.
To specify the sensing power of the many-body interacting probe, in 
Fig.~\ref{fig:nonLinearEffect_MB} (a) we present the QFI as a function of $h$ for system of sizes $L{\in}\{4,\cdots,16\}$ under the effect of $V_i{=}hi^2$.
After a size-dependent plateau in the ergodic phase, one observes a rapid decay of the QFI in the localized phase.  
The universal decrease of the QFI in the localized phase, described by $\mathcal{F}_{Q}{\propto}|h-h^{\max}|^{-\alpha}$ with $\alpha{\simeq}4$ has been shown by the dashed line. Note that, our analysis leads to $\alpha{\simeq}4$ for all the studied values of $\gamma$'s. 
In many-body interacting probe, the reduction of the Hamiltonian Eq.~(\ref{Eq:Stark_MB_Hamiltonian}) to its spectrum is not trivial.
In this case, one can only rely on the numerical study to elucidate the physics of the system.
By enlarging the system a pronouncing peak in the QFI appears, see Fig.~\ref{fig:nonLinearEffect_MB} (a), indicating the transition point. However, to have a consistence analysis based on all the considered $L$s, we focus on the performance in the extended phase, namely $h{=}10^{-6}J$.  
The value of QFI, offers a scaling behavior as $\mathcal{F}_{Q}(h{=}10^{-6}J,\gamma){\propto}L^{\beta(\gamma)}$, see the inset of Fig.~\ref{fig:nonLinearEffect_MB} (b). The extracted $\beta$'s as a function of $\gamma$, is reported in Fig.~\ref{fig:nonLinearEffect_MB} (b). 
The numerical simulations are well-described by Eq.~(\ref{Eq.Beta})
with $(a,b){\sim}(3.69,-0.45)$. Here, results are obtained from analyzing many-body interacting systems of sizes $L{\in}\{6,\cdots,16\}$.
\\

Several important observation need to be highlighted.
First, by increasing nonlinearity, the super-Heisenberg scaling precision of both Stark probes improves remarkably.
The origin of this improvement is the enhancement in the off-resonant energy splitting between neighboring sites. 
While in $V_i{=}hi$ the energy splitting between consecutive  sites is $h$, in $V_i{=}hi^2$ one has $(2i{+}1)h$.
This increase in the off-resonant energy splitting not only elevates the distinguishability of the energy difference but also boosts the power of localization in a way that the transition point vanishes even in finite-size systems. 
Second, in single-particle probes, quantum-enhanced sensitivity can be obtained for all values of $\gamma$ while in many-body probes it can only be achieved for $\gamma{>0.5}$.
Third, the nonlinearity of the gradient field plays stronger role in the many-body probe and results in sharper growth of $\beta$ in comparison with the single-particle probe. This hints that for $\gamma>2.5$, a many-body probe operates better for reasonably large system sizes.
\\

It is worth emphasizing that in the extended phase the superposition of different excitations $|j\rangle$ extends over a long distance, see Eq.~(\ref{Eq:eigensystem}), making the quantumness of the probe very pronounced. This manifestation of quantumness is indeed the feature which is responsible for creating super-linear scaling of the QFI in the extended phase. On the other hand, in the localized phase the extension of the superposition is compressed to a few neighboring sites, reducing the quantumness of the probe.  In this case, an excitation may be delocalized over a finite number of sites, independent of $L$. In such systems, even in the presence of $N$ excitations, the performance of the probe can be approximated as $N$ identical probes each containing one excitation delocalized over a short fixed length which is independent of $L$. In this case, the QFI of the system becomes $\mathcal{F}_Q(h){\propto}N$~\cite{manshouri2024quantum}. 
Even in the half-filled probe with $N{=}L/2$ excitations the QFI scales linearly with system size $L$, resembling the standard limit. In the case of fixed excitation $N$, the QFI becomes size independent, as demonstrated in Fig.~\ref{fig:nonLinearEffect_SP}(a), consistent with the results in Ref.~\cite{manshouri2024quantum}.  
\\

\section{Parabolic Gradient Field Sensing:\\ Multi-Parameter Estimation}\label{S.NSP-MPE}
\begin{figure*}[t]
\includegraphics[width=\linewidth]{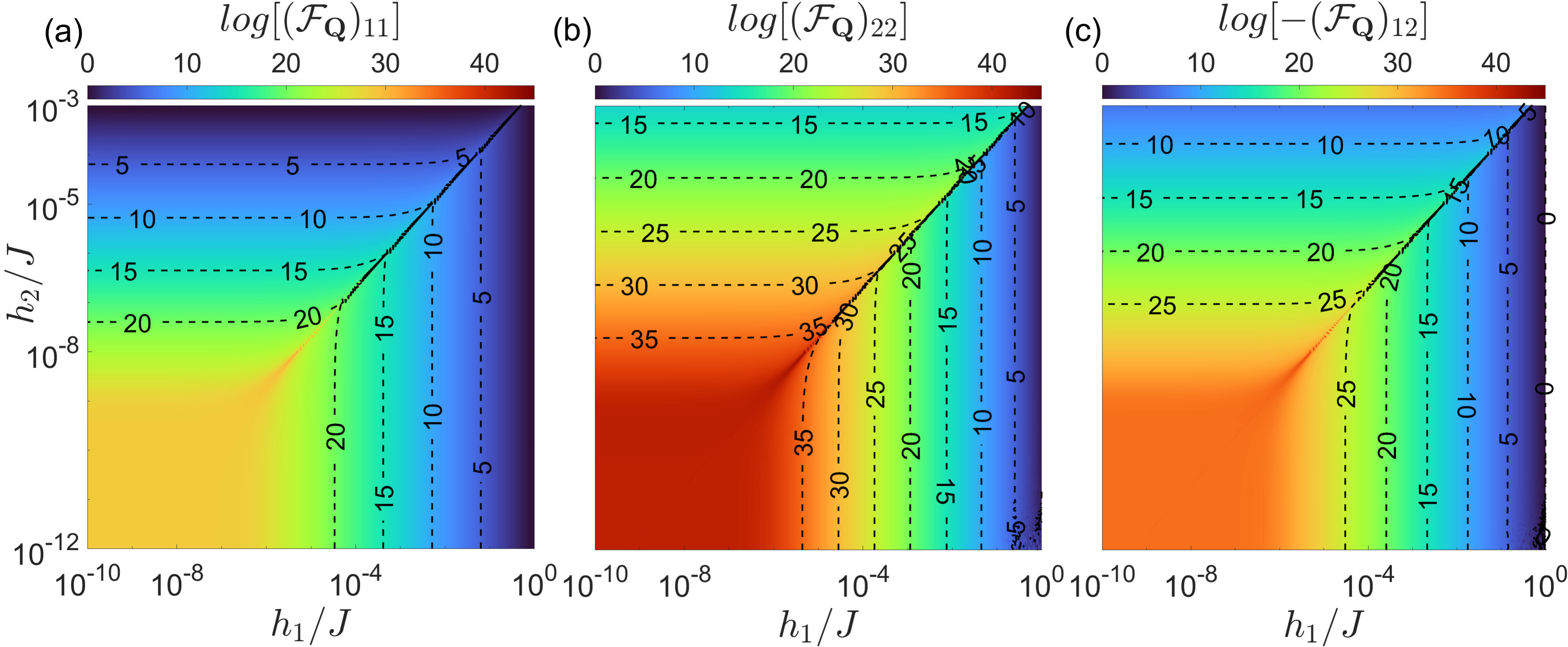}
\includegraphics[width=\linewidth]{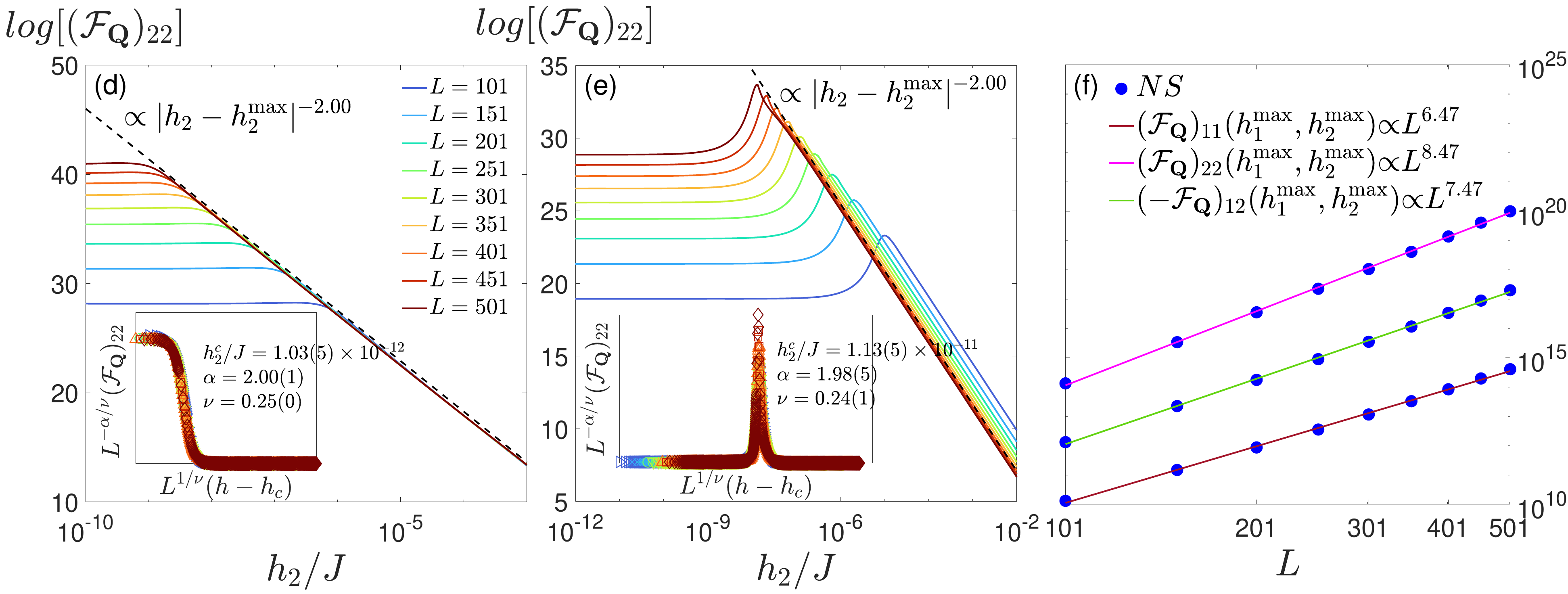}
\caption{The QFI matrix elements, (a) ${\rm log}[(\boldsymbol{\mathcal{F}_{Q}})_{11}]$, (b) ${\rm log}[(\boldsymbol{\mathcal{F}_{Q}})_{22}]$ and (c) ${\rm log}[(-\boldsymbol{\mathcal{F}_{Q}})_{12}]$, as a function of linear $h_{1}$ and nonlinear $h_{2}$ terms of the potential landscape $V_i$, when the probe with size $L=501$ is prepared in the ground state.   ${\rm log}[(\boldsymbol{\mathcal{F}_{Q}})_{22}]$ as a function of $h_2$ for a fixed value of (d) $h_1/J=10^{-8}$ and (e) $h_1=h_2(L-1)$. The dashed lines describe the algebraic decay of the Fisher information in the localized phase which is well-fitted by $|h_{2}-h_{2}^{\max}|^{-\alpha}$ with $\alpha{=}2.00$. 
The insets are the corresponding finite-size scaling analysis. The optimal data collapse is obtained for the reported $(h_2^{c},\alpha,\nu)$.
Note that all the axes are in a logarithmic scale. 
(f) The maximum values of the QFI matrix elements (dots) as a function of probe size $L$.
Lines are the best fitting function as $(\boldsymbol{\mathcal{F}_{Q}})_{ij}(h_1^{\max},h_2^{\max})\propto L^{\beta_{ij}}$  to describe the scaling behavior of the QFI matrix elements.
}\label{fig:QFIelementsSP}
\end{figure*}
\begin{figure}[t]
\includegraphics[width=\linewidth]{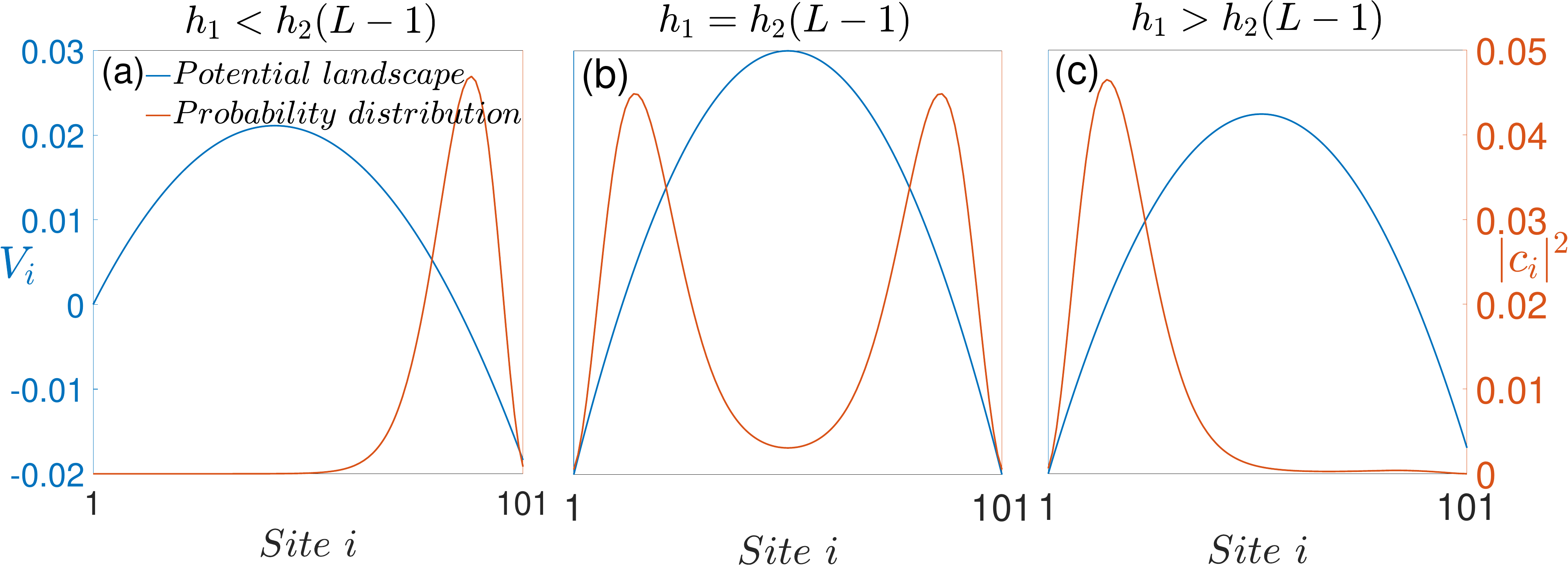}
\caption{The potential landscape $V_i{=}h_1(i-1){-}h_2(i-1)^2$ and probability distribution of the ground state $|c_i|^2$, labeled on the left and right side of the plots, respectively, as a function of lattice site $i$ with probe size $L{=}101$. The values of the parameters are chosen as (a) $h_1{<}h_2(L-1)$, (b) $h_1{=}h_2(L-1)$, and (c) $h_1{>}h_2(L-1)$.
}\label{fig:PotentialProbabilityGS}
\end{figure}
Having elucidated the sensing power of the Stark probes in estimating nonlinear gradient field $V_i{=}hi^\gamma$ with only one parameter under scrutiny, namely $h$, in this section we aim to expand our study to a wider range of potential forms.
Relying on multi-parameter estimation theory, in this section we assess the performance of the Stark probes for estimating a parabolic potential landscape as 
\begin{equation}\label{Eq:Vi}
V_{i}(h_1,h_2) = h_1(i-1)-h_2(i-1)^2.
\end{equation}
Note that linear, $h_1$, and nonlinear, $h_2$, gradient fields in $V_i$ that are under estimation, compete to localize the system through their opposite singes.

\subsection{Single-particle probe}\label{SubS.NMP-SPP}
We start with the Hamiltonian Eq.~(\ref{Eq:Stark_Hamiltonian}) in which gradient potential $V_i$ is replaced by Eq.~(\ref{Eq:Vi}). 
The aim is to estimate both parameters $h_1$ and $h_2$ when this probe is initialized in its ground state. 
As the figure of merit, we compute the QFI matrix whose elements are plotted in Fig.~\ref{fig:QFIelementsSP}. 
Since the QFI matrix is symmetric, one has  $(\boldsymbol{\mathcal{F}_{Q}})_{12}{=}(\boldsymbol{\mathcal{F}_{Q}})_{21}$. 
In Figs.~\ref{fig:QFIelementsSP}(a)-(c), we depict $(\boldsymbol{\mathcal{F}_{Q}})_{11}$, $(\boldsymbol{\mathcal{F}_{Q}})_{22}$ and $(\boldsymbol{\mathcal{F}_{Q}})_{12}$ as a function of $h_1$ and $h_2$ for the probe of size $L{=}501$, respectively. 
The phase diagram is indeed fully described by both $h_1$ and $h_2$.
Several common features can be observed.  
By changing the parameters $(h_1,h_2)$, all the elements of the QFI matrix show a clear transition from the extended phase in which the Fisher information remains steadily high for small values of $h_1$ and $h_2$ (a rectangular region in the lower left corner of the phase diagram with hot color) to a localized phase in which the Fisher information significantly shrinks (regimes with cold colors).
The Fisher information reveals a peak along the line $h_1{=}h_2(L-1)$ that represents the situation in which the Stark potential $V_i$ becomes symmetric around the center of the system, namely $V_i{=}V_{L-i+1}$. 
We come back to this interesting case later. 
To clarify this Stark localization transition in our probe, in Fig.~\ref{fig:QFIelementsSP}(d) we plot $(\boldsymbol{\mathcal{F}_{Q}})_{22}$ versus $h_2$ for $h_1{=}10^{-8}J$ and different probe sizes.
The QFI initially follows a plateau, indicating the extended phase, and then starts to decrease at a specific value of $h_2=h_2^{\max}$, which is size dependent.
By increasing the size of the system, three important features can be observed. 
First, $h_2^{\max}$'s tend to smaller values signaling  $h_2^{\max}{\rightarrow}h_2^{c}{=}0$ in the thermodynamic limit.
Second, the value of the QFIs in the extended phase dramatically increases by enlarging the system size, hinting its divergence in the thermodynamic limit (i.e.  $L{\rightarrow}\infty$).
Third, in the localized phase the QFIs become size independent and show a universal algebraic decay as ${\propto}|h_2 - h_2^{\max}|^{-\alpha}$ with $\alpha{=}2.00$, see the dashed fitting line in the panel (d).
Note that these three observations are valid for all the elements of the QFI matrix (data not shown). Similarly, one can fix $h_2$ into a small value and plot the elements of the QFI matrix as a function of $h_1$ which all show similar qualitative behavior, namely a size-dependent plateau followed by a universal size-independent behavior. 
This is analogous to the Stark localization transition observed for a single parameter~\cite{he2023stark,Yousefjani2023}.
In Fig.~\ref{fig:QFIelementsSP}(e), we investigate the behavior of the QFI along the symmetric line $h_1{=}h_2(L-1)$ for various system sizes. 
Interestingly, $(\boldsymbol{\mathcal{F}_{Q}})_{22}$ shows a peak at the transition from the extended to the localized phase. 
The emergence of peaks during the symmetric line can be observed in all the QFI matrix elements (shown as dark points on Figs.~\ref{fig:QFIelementsSP}(a)-(c)).
\\

To elucidate the origin of this extra enhancement in the QFI elements, we note that for $h_1{=}h_2(L{-}1)$, one has $[H,\mathcal{M}]{=}0$, where $\mathcal{M}$ is the mirror operator defined as $\mathcal{M}|i_1,\cdots,i_L\rangle{=}|i_L,\cdots,i_1\rangle$. 
This implies that the eigenstates of the system are either symmetric, namely $\mathcal{M}|E_k\rangle{=}|E_k\rangle$, or anti-symmetric, namely $\mathcal{M}|E_k\rangle{=}{-}|E_k\rangle$, around the center of the chain.
We conclude that the presence of this mirror symmetry is the origin of the extra sensitivity that one observes in Figs.~\ref{fig:QFIelementsSP}(e).
For the sake of completeness, we also investigate the wave function of the ground state $|E_1\rangle{=}\sum_i c_i |i\rangle$ for arbitrary values of $V_i$. 
In Figs.~\ref{fig:PotentialProbabilityGS}(a)-(c), both $V_{i}$ and $|c_i|^2$ as a function of lattice site $i$ and in three different regimes are presented.
The mirror symmetry of the ground state for $h_1{=}(L-1)h_2$, see Fig.~\ref{fig:PotentialProbabilityGS}(b), results in the bilocalization of the particle in both edges of the system.
By getting distance from the symmetric line, for instance in regimes with $h_1{<}(L-1)h_2$ or $h_1{>}(L-1)h_2$, the mirror symmetry breaks and, hence,  $[H,\mathcal{M}]{\neq}0$, therefore the particle fully localize in either left or right side of the chain, see Figs.~\ref{fig:PotentialProbabilityGS}(a) and (c).
\\

To determine the quantum enhancement in terms of the system size, in Fig.~\ref{fig:QFIelementsSP} (f) we plot the maximum values of the QFI matrix elements along the symmetric line 
$h_1{=}h_2(L-1)$ as a function of the probe size.
Clearly, the numerical simulations (NS) are well defined by a fitting function of the form $(\mathcal{F}_{Q})_{ij}{}\propto L^{\beta_{ij}}$ (solid lines).
Interestingly all the QFI matrix elements provide super-Heisenberg scaling as 
$(\boldsymbol{\mathcal{F}_{Q}})_{11}{\propto}L^{6.47}$, 
$(\boldsymbol{\mathcal{F}_{Q}})_{22}{\propto}L^{8.47}$,
and 
$(-\boldsymbol{\mathcal{F}_{Q}})_{12}{\propto}L^{7.47}$.
To describe the Stark transition, we rely on the continuous phase transition framework which suggests that the QFI matrix elements satisfy the following ansatz
\begin{equation}
(\boldsymbol{\mathcal{F}_{Q}})_{ii}=L^{\alpha_i/\nu_i} \mathcal{G}_{i}(L^{1/\nu_i} (h-h_c)) 
\end{equation}
where $\mathcal{G}_{i}(\bullet)$ is an arbitrary function and $\alpha_i$ and $\nu_i$ are critical exponents. 
One can extract the critical exponents through finite-size scaling analysis in which the quantity $L^{-\alpha_i/\nu_i}(\boldsymbol{\mathcal{F}_{Q}})_{ii}$ is plotted versus $L^{1/\nu_i} (h{-}h_c)$ for various system sizes. 
By varying the critical exponents, one can collapse the curves of different sizes. 
In the inset of Fig.~\ref{fig:QFIelementsSP} (d), the corresponding data collapse for $(\mathcal{F}_{Q})_{22}$ is obtained for $(h_2^c,\alpha_2,\nu_2){=}(1.03{\times}10^{-12}J,2.00,0.25)$. 
Similarly, for the transition along the symmetry line, the data collapse shown in the inset of  Fig.~\ref{fig:QFIelementsSP} (e) is obtained for 
$(h_2^c,\alpha_2,\nu_2){=}(1.10{\times}10^{-11}J,1.98,0.24)$.
Applying the finite-size scaling analysis for $(\mathcal{F}_{Q})_{11}$ results in $(h_1^c,\alpha_1,\nu_1){=}(1.14{\times}10^{-10}J,1.99,0.31)$ and $(h_1^c,\alpha_1,\nu_1){=}(1.02{\times}10^{-10}J,1.99,0.33)$ for the transition during the symmetric line and beyond it.   
For the obtained critical parameters, one can check the validity of $\beta_i{=}\alpha_i/\nu_i$ (for $i{=}1,2$) which shows that the critical exponents are not independent of each other, see Ref.~\cite{he2023stark} for more details.
%%%%%%%%%%%%%%%%%%%%%%%%%%%%%%%%%%%%%%%%%%%%%%%%%%
\subsection{Many-body interacting probe}\label{SubS.NMP-MBP}
\begin{figure*}[t]
\includegraphics[width=0.49\linewidth]{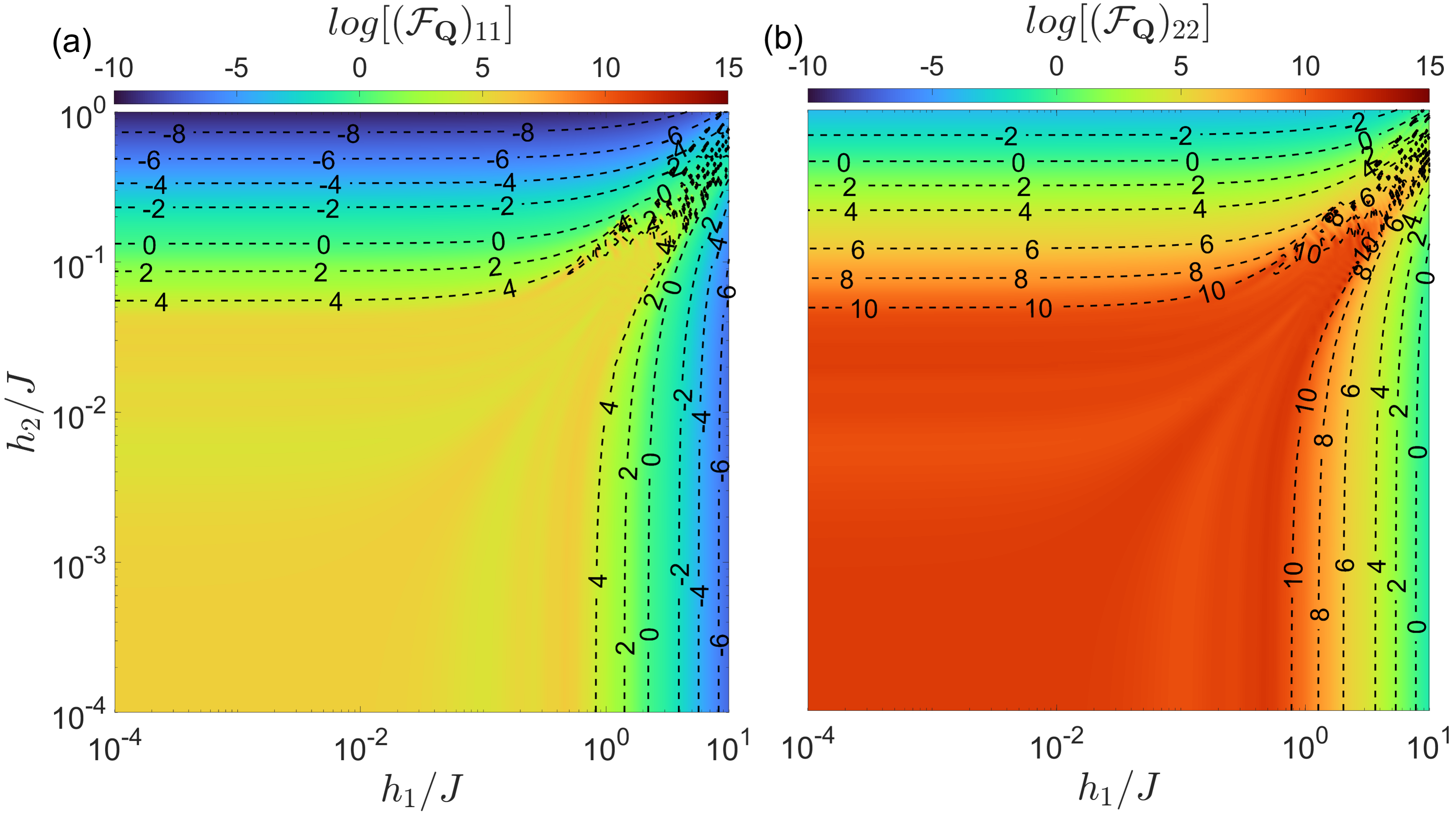}
\includegraphics[width=0.49\linewidth]{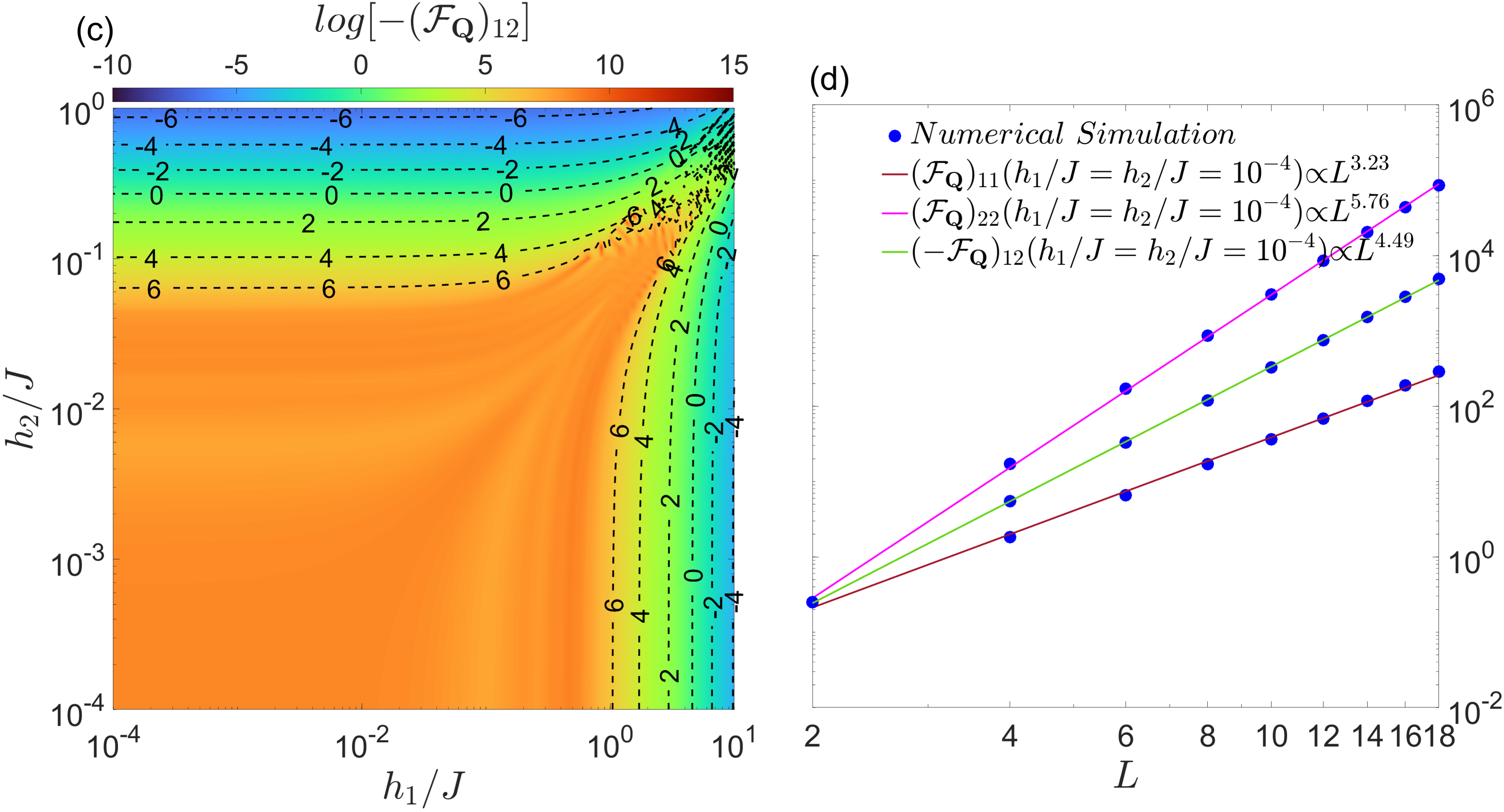}
\caption{ The QFI matrix elements, (a) ${\rm log}[(\boldsymbol{\mathcal{F}_{Q}})_{11}]$ and (b) ${\rm log}[(\boldsymbol{\mathcal{F}_{Q}})_{22}]$, as a function of $h_{1}$ and $h_{2}$ when the many-body interacting probe (Eq.~(\ref{Eq:Stark_MB_Hamiltonian})) with size $L=18$ is prepared in its ground state. (d) the values of the QFI matrix elements at $(h_1{=}h_2{=}10^{-4}J)$ (deep in the delocalized phase) as a function of the probe size. The numerical results are properly described by a fitting function as $(\boldsymbol{\mathcal{F}_{Q}})_{ij}(h_1,h_2)\propto L^{\beta}$ with reported $\beta$'s.}\label{fig:QFIelementsMB}
\end{figure*}
In contrast to interferometry-based quantum sensing in which interaction between particles deteriorates the sensitivity, strongly correlated many-body probes exploit the interaction to enhance their sensitivity.
Therefore, studying the impact of interactions on our Stark probe's performance is valuable.
We start by calculating the QFI matrix elements for the ground state of systems up to size $L{=}18$, obtained using exact diagonalization.
The results are presented in Figs.~\ref{fig:QFIelementsMB}(a)-(c).
Similar to the single-particle probe, by varying $h_1$ and $h_2$ from small to large values, the QFI matrix elements change remarkably from a region where their value remains steady (the area with warm colors) in the delocalized phase to a region where their values decrease monotonically (the area with cold colors) in the localized phase.
Although the finite-size effect in many-body interacting probes results in a wider area for the extended phase, by increasing the size of the system this area shrinks until eventually vanishes at the thermodynamic limit. 
Considering the strong finite-size effect on the results, extracting the scaling behavior in the vicinity of the phase boundaries is very challenging.
Therefore, in Fig.~\ref{fig:QFIelementsMB}(d), we focus on the delocalized phase and plot the value of QFI matrix elements at $h_1{=}h_2{=}10^{-4}J$.
The numerical results (markers) are well-describe by the fitting function as $(\boldsymbol{\mathcal{F}_{Q}})_{ij}\propto L^{\beta}$ (solid line) with $\beta{>}2$ for all QFI matrix elements.
The exact values for $h_1{=}h_2{=}10^{-4}J$ are obtained as  
$(\boldsymbol{\mathcal{F}_{Q}})_{11}{\propto}L^{3.23}$, 
$(\boldsymbol{\mathcal{F}_{Q}})_{22}{\propto}L^{5.76}$,
and 
$(-\boldsymbol{\mathcal{F}_{Q}})_{12}{\propto}L^{4.49}$.
These results guarantee that similar to the single-particle probe, the many-body interacting probe can also offer quantum-enhanced sensitivity.
%%%%%%%%%%%%%%%%%%%%%%%%%%%%%%%%%%%%%%%%%%
\subsection{Optimal measurement}\label{SubS.MPE-OM}
\begin{figure}[t]
    \centering
    \includegraphics[width=\linewidth]{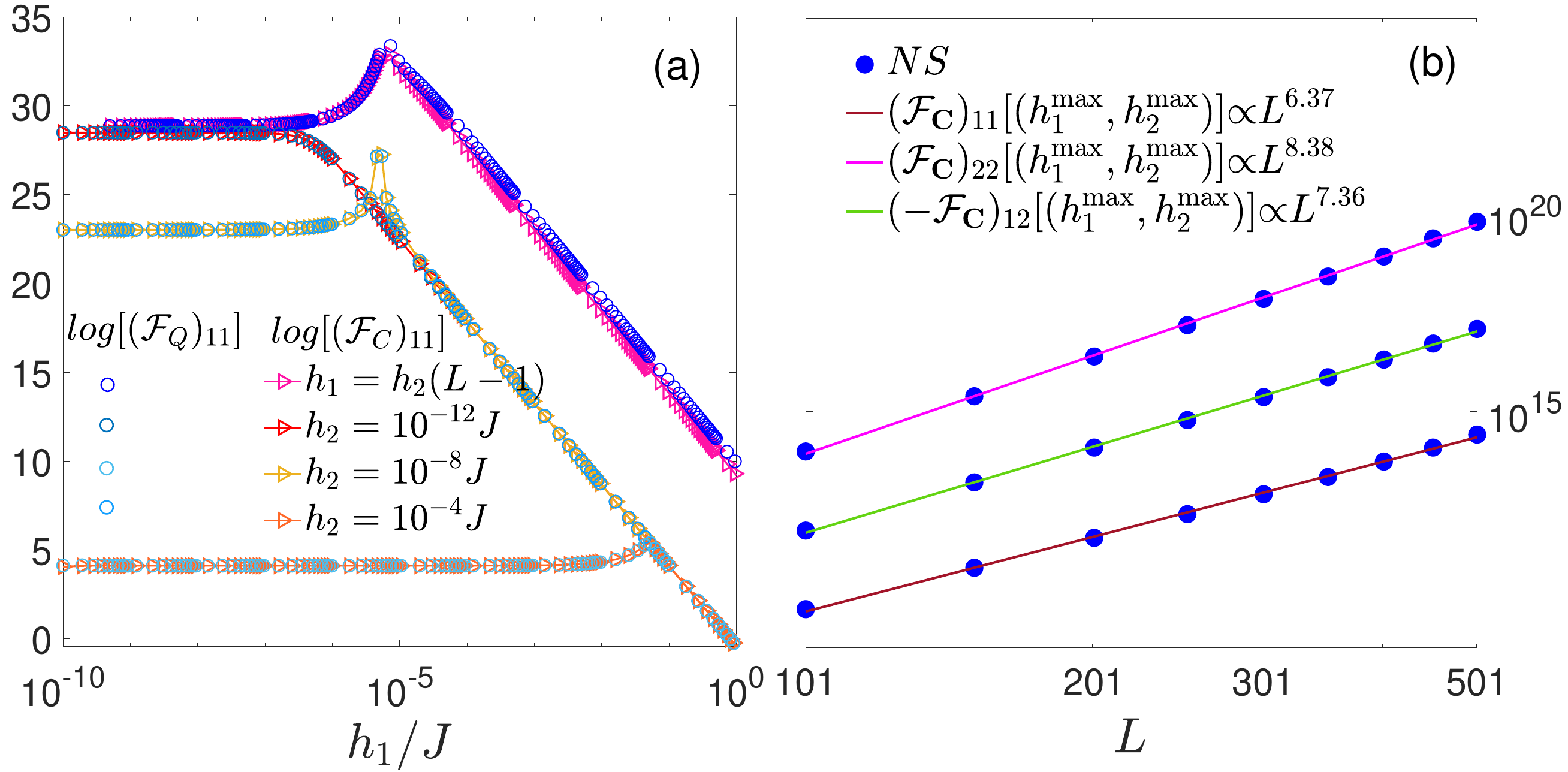}
    \includegraphics[width=\linewidth]{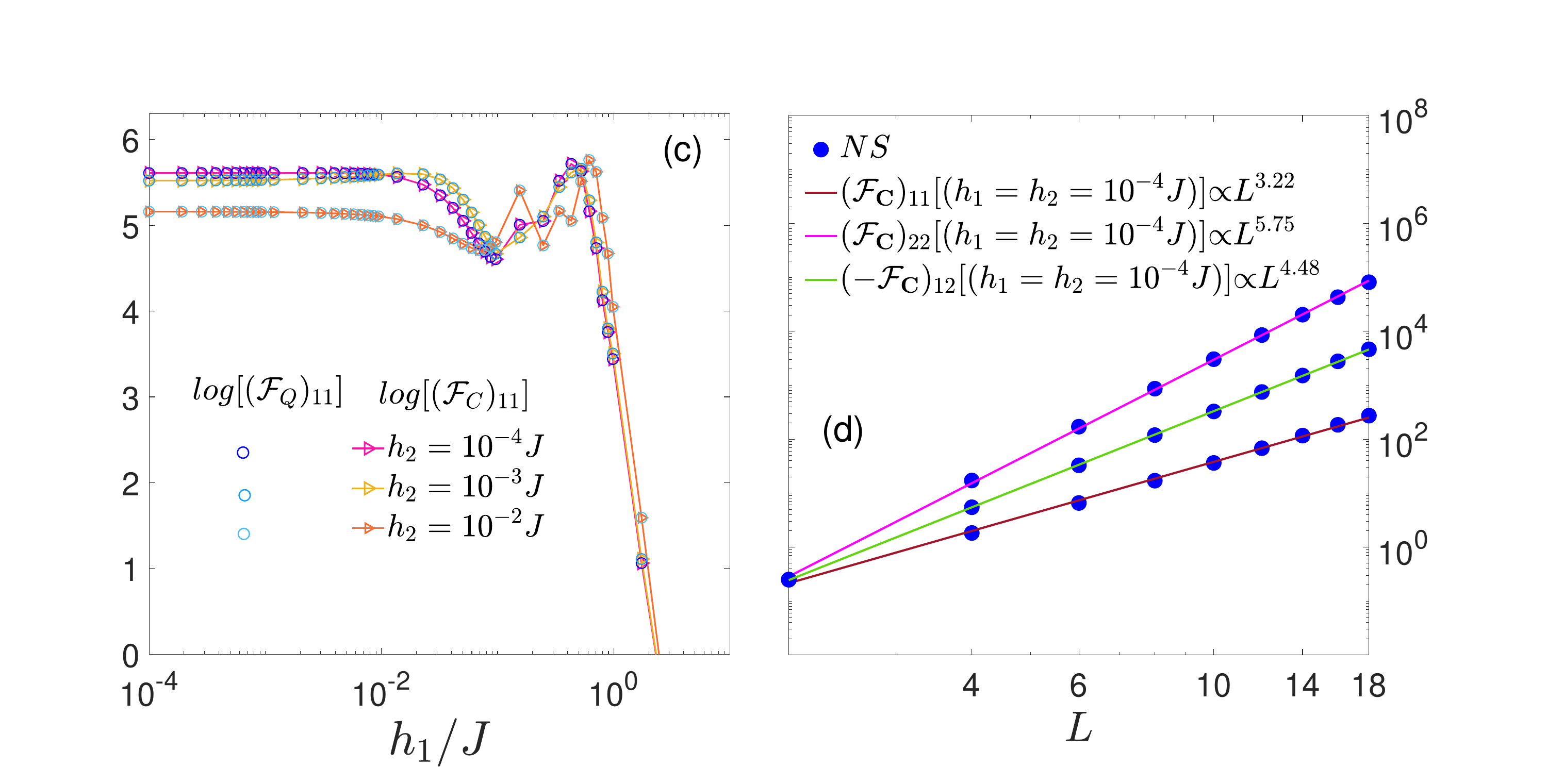}
    \caption{ $(\mathcal{F}_{C})_{11}$ and $(\mathcal{F}_{Q})_{11}$ versus $h_1$ in various $h_2$ for (a) single-particle and (c) many-body interacting probe. 
    The maximum values of the CFI matrix elements for (b) single-particle and (d) many-body interacting probes versus $L$. The scaling behavior of all the matrix elements is obtained as $(\boldsymbol{\mathcal{F}_{C}})_{ij}(h_1^{\max},h_2^{\max})\propto L^{\beta_{ij}}$ with reported $\beta$s. }
    \label{fig:CFI}
\end{figure}
Capturing the multi-parameter quantum Cram\'{e}r-Rao inequality in Eq.~(\ref{Eq.QCRBM}) demands a POVM that is optimized concerning all unknown parameters.
As has been discussed before,  this relies on the satisfaction of either $[L_{i},L_{j}]{=}0$ or $\text{Tr}(\rho({\boldsymbol{h}})[L_{i},L_{j}]){=}0$.
One of the striking properties of our model is that, while SLD operators do not satisfy the former except for $h_1{=}(L-1)h_2$, they always satisfy the latter one.
This implies that there is always a set of optimized measurements that guarantee the availability of the quantum multi-parameter Cram\'{e}r-Rao bound.
In the single-particle level, we propose a simple position measurement described by the POVM $\{\Pi_k{=}|k\rangle \langle k|\}_{k=1}^L$.
Interestingly, the CFI matrix elements in Eq.~(\ref{Eq.CCRBM_elements}) highly resemble their quantum counterpart. This is evidenced in Fig.~\ref{fig:CFI} (a) wherein $(\mathcal{F}_C)_{11}$ and $(\mathcal{F}_Q)_{11}$ as a function of $h_1$ and for various $h_2$, including $h_2{=}h_1(L-1)$, are plotted as an illustrative example.
Regarding the scaling behavior, in Fig.~\ref{fig:CFI} (b), we report the maximum values of the CFI matrix elements, happening along the symmetric line $h_1{=}h_2(L-1)$, as a function of $L$.
The markers are the numerical results for various sizes, which are well described by the fitting function $(\boldsymbol{\mathcal{F}_{C}}(\boldsymbol{h}))_{ij}{\propto}L^{\beta_{ij}}$.
The obtained exponents are $(\boldsymbol{\mathcal{F}_{C}}(\boldsymbol{h}))_{11}{\propto}L^{6.37}$, $(\boldsymbol{\mathcal{F}_{C}}(\boldsymbol{h}))_{22}{\propto}L^{8.38}$, and $(-\boldsymbol{\mathcal{F}_{C}}(\boldsymbol{h}))_{12}{\propto}L^{7.36}$. 
\\

Similar to the single-particle probe, the many-body interacting stark probes satisfy ${\rm Tr}(\rho(\boldsymbol{h})[L_i,L_j]){=}0$. This hints that there are a set of optimal POVM that can provide Eq.~(\ref{Eq.QCRBM}). We find that a set composed of all possible spin configurations in the half-filling sector results in $(\boldsymbol{\mathcal{F}_{C}})_{ij}(\boldsymbol{h}) {\simeq} (\boldsymbol{\mathcal{F}_{Q}})_{ij}(\boldsymbol{h})$, see Fig.~\ref{fig:CFI} (c) that, for instance, report both $(\mathcal{F}_C)_{11}$ and $(\mathcal{F}_Q)_{11}$ as function of $h_1$ and various $h_2$.
Surprisingly the scaling behavior of the CFI matrix elements resembles those of QFI matrix elements, see Fig.~\ref{fig:CFI} (b). Our analysis led to   
$(\boldsymbol{\mathcal{F}_{C}})_{11}{\propto}L^{3.37}$, 
$(\boldsymbol{\mathcal{F}_{C}})_{22}{\propto}L^{5.66}$,
and 
$(-\boldsymbol{\mathcal{F}_{C}})_{12}{\propto}L^{4.52}$.
All the above analysis shows that the super-Heisenberg scaling precision offered by the Stark probes is, indeed, obtainable using experimentally available measurements.

\subsection{Resource analysis}\label{SubS.MPE-RA}
\begin{figure}[t]
\includegraphics[width=\linewidth]{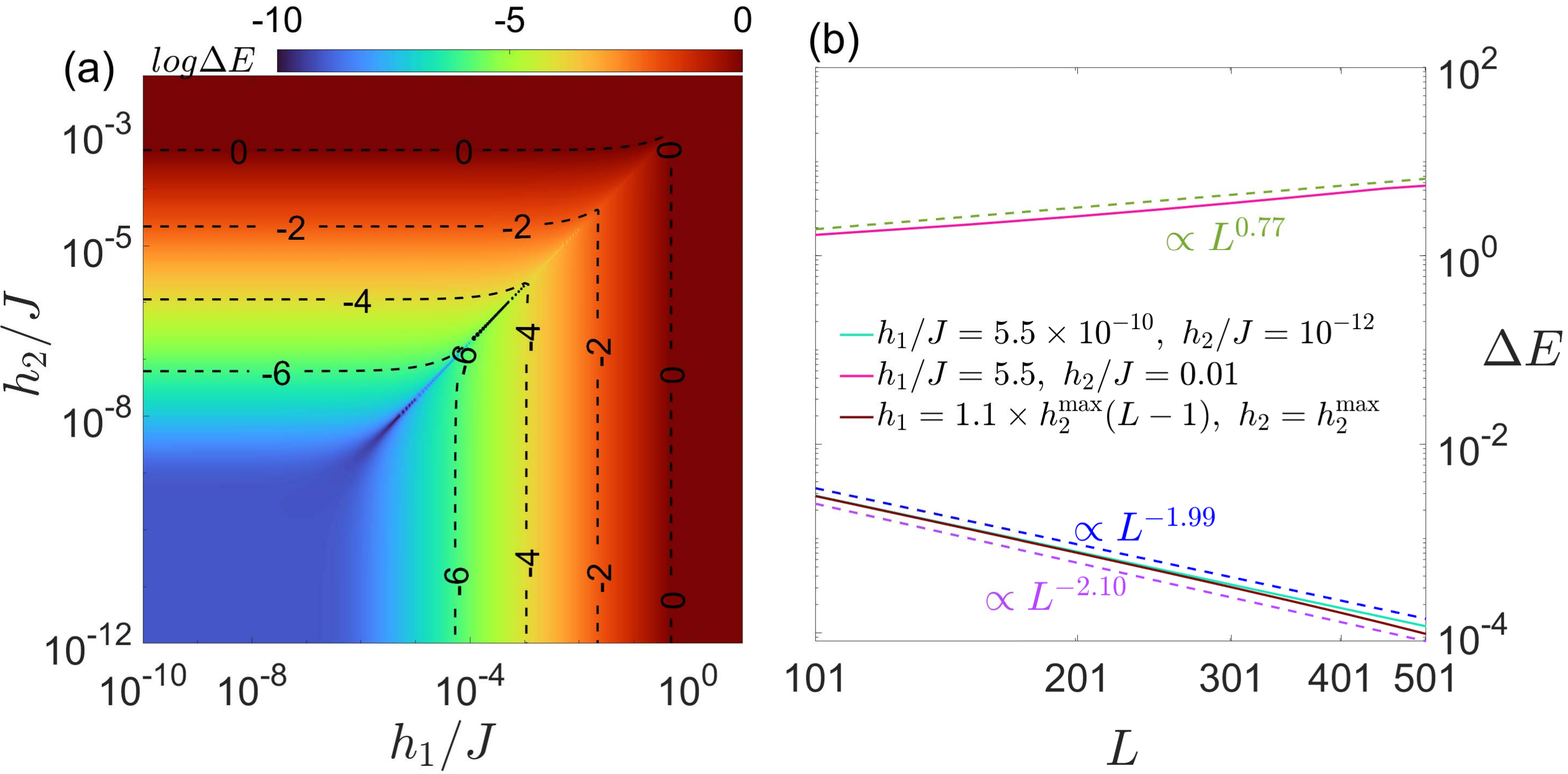}
\includegraphics[width=\linewidth]{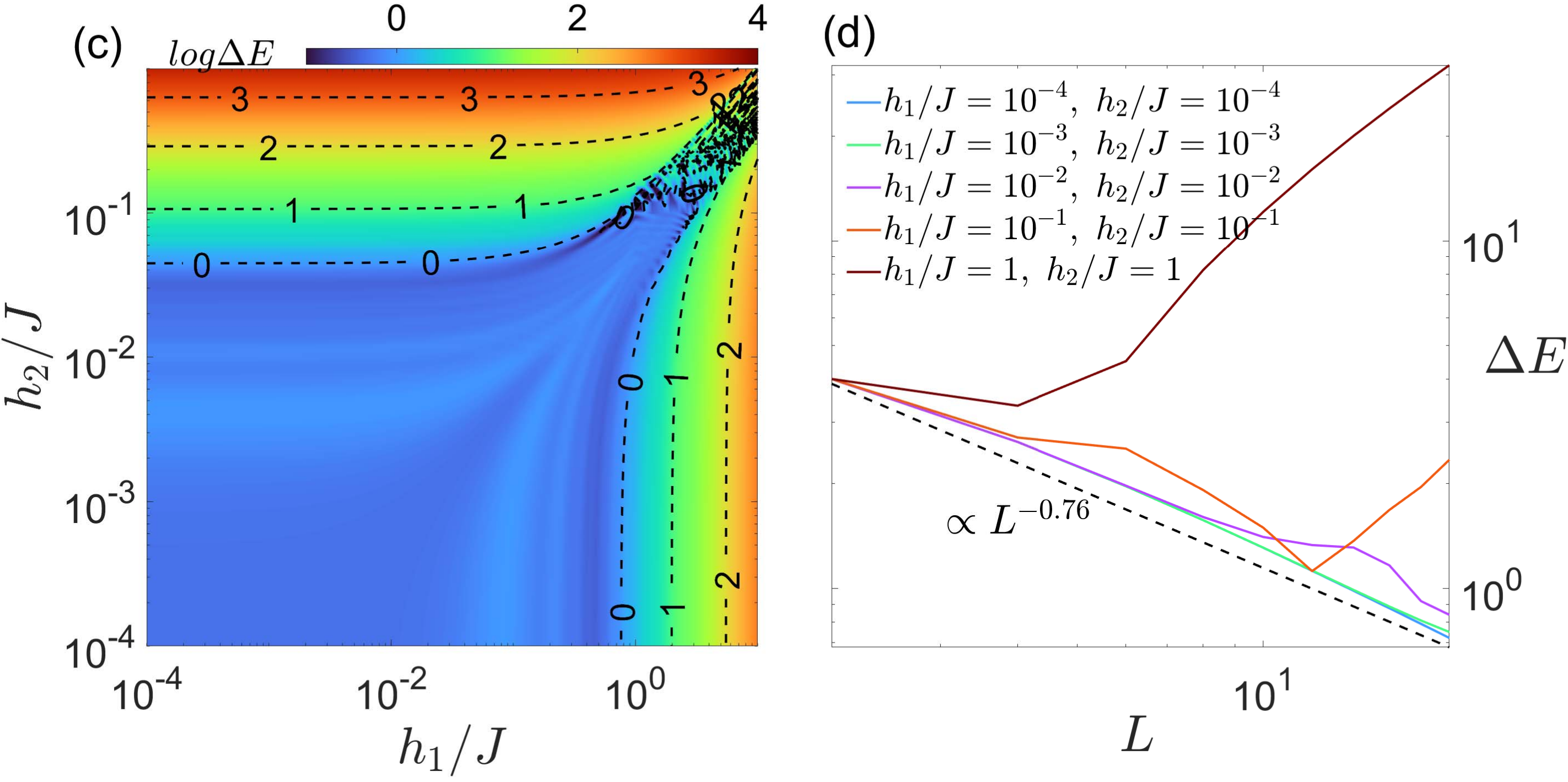}
\caption{The energy gap between the ground state and the first-excited state $\Delta E$ as a function of $h_1$ and $h_2$ for (a) single-particle probe with size $L{=}501$ and (c) many-body interacting probe of size $L{=}20$. The energy gap as a function of probe size $L$ for different choices of $h_1$ and $h_2$ in (b) single-particle and (d) many-body interacting probes. The dashed lines are eye guides that describe the decreasing behavior as $\Delta E{\sim}L^{-z}$. 
}\label{fig:DE}
\end{figure}
As discussed above, the ground state of Stark probes can achieve quantum-enhanced sensitivity. So far, in our analysis, we have considered the probe size $L$ as the relevant resource for achieving such enhanced precision. However, preparation of the ground state might be challenging and time-consuming. Therefore, one may include the time $t$ which is needed to prepare the probe in its ground state as another resource.
In order to incorporate time into our resource analysis, we use the normalized QFI matrix  $t^{-1}\boldsymbol{\mathcal{F}_{Q}}(\boldsymbol{h})$ as a figure of merit. 
To estimate the time $t$, we consider adiabatic state preparation in which one can slowly evolve the probe from a simple ground state into the desired one.
To avoid the emergence of excited states during this evolution and guarantee that the system ends up in the desired ground state, 
the quench rate of the parameters needs to be adequately small, namely $t\sim 1/\Delta E$ with $\Delta E$ being the lowest energy gap during the variation of the Hamiltonian~\cite{Teufel2022}. 
When the system is adiabatically evolved near a phase transition point, the minimum energy gap at the criticality scales as  $\Delta E{\sim}L^{-z}$~\cite{rams2018limits}, where $z$ is known as the dynamical critical exponent. Therefore, the time required for state preparation is $t\sim L^z$.  
In Fig.~\ref{fig:DE}(a), we plot the energy gap between the ground state and the first excited state $\Delta E$ as a function of $h_1$ and $h_2$ for the single-particle probe of size $L{=}501$.
By moving from the extended phase to the localized one, the energy gap $\Delta E$ increases.
Fig.~\ref{fig:DE}(b) illustrates the obtained $\Delta E$ for different sizes of the probe, in three points including deep in the extended phase, namely $h_{1}{=}5.5\times 10^{-10}J$ and $h_{2}{=}10^{-12}J$, near the transition point ($h_1^{\max},h_2^{\max}$), and deep in the localized side, namely $h_{1}{=}5.5J$ and $h_{2}{=}0.01J$.
Not that to avoid the ground-state degeneracy in the transition point ($h_1^{\max},h_2^{\max}$) which is across the symmetric line  $h_1=h_2(L-1)$, we focus on 
$h_1=1.1\times h_2(L-1)$.
In both the extended phase and near the transition points, one has $\Delta E{\propto}L^{-1.99}$ and $\Delta E{\propto}L^{-2.10}$, respectively, which is in agreement with our previous observations~\cite{he2023stark}. 
However, the energy gap becomes positively correlated to the probe size as $\Delta E{\propto}L^{0.77}$ in the localized phase, which might be beneficial for the state initialization.
Based on this result the ultimate scaling of the QFI matrix elements is obtained as $t^{-1}(\boldsymbol{\mathcal{F}_{Q}})_{11}(h_1^{\max},h_2^{\max}){\propto}L^{4.37}$, 
$t^{-1}(\boldsymbol{\mathcal{F}_{Q}})_{22}(h_1^{\max},h_2^{\max}){\propto}L^{6.37}$,
and 
$t^{-1}(-\boldsymbol{\mathcal{F}_{Q}})_{12}(h_1^{\max},h_2^{\max}){\propto}L^{5.37}$, confirming the quantum-enhancement in the achievable precision.
\\
Following the same resource analysis for the single-particle probe, here, we characterize the final scaling behavior of the QFI matrix elements, after considering the preparation time. 
In Fig.~\ref{fig:DE}(c), the energy gap $\Delta E$ between the ground state and the first excited state as a function of $h_1$ and $h_2$ for a probe of size $L{=}20$ is reported.
Obviously, increasing the parameters widens the energy gap.
Extracting the dynamical critical exponent through studying $\Delta E$ versus $L$ in Fig.~\ref{fig:DE}(d) results in $z{\sim}0.76$ for a many-body probe that works deeply in the delocalized phase.
This results in the following normalization for the QFI matrix elements as
$t^{-1}(\boldsymbol{\mathcal{F}_{Q}})_{11}(h_1{=}h_2{=}10^{-4}J){\propto}L^{2.47}$, 
$t^{-1}(\boldsymbol{\mathcal{F}_{Q}})_{22}(h_1{=}h_2{=}10^{-4}J){\propto}L^{5.00}$,
and 
$t^{-1}(-\boldsymbol{\mathcal{F}_{Q}})_{12}(h_1{=}h_2{=}10^{-4}J){\propto}L^{3.73}$.
Indeed, the quantum enhancement can still be achieved even after considering the time that one needs to spend for initializing the probe.

\subsection{Simultaneous estimation}\label{SubS.MPE-SE}
\begin{figure}[t]
\includegraphics[width=\linewidth]{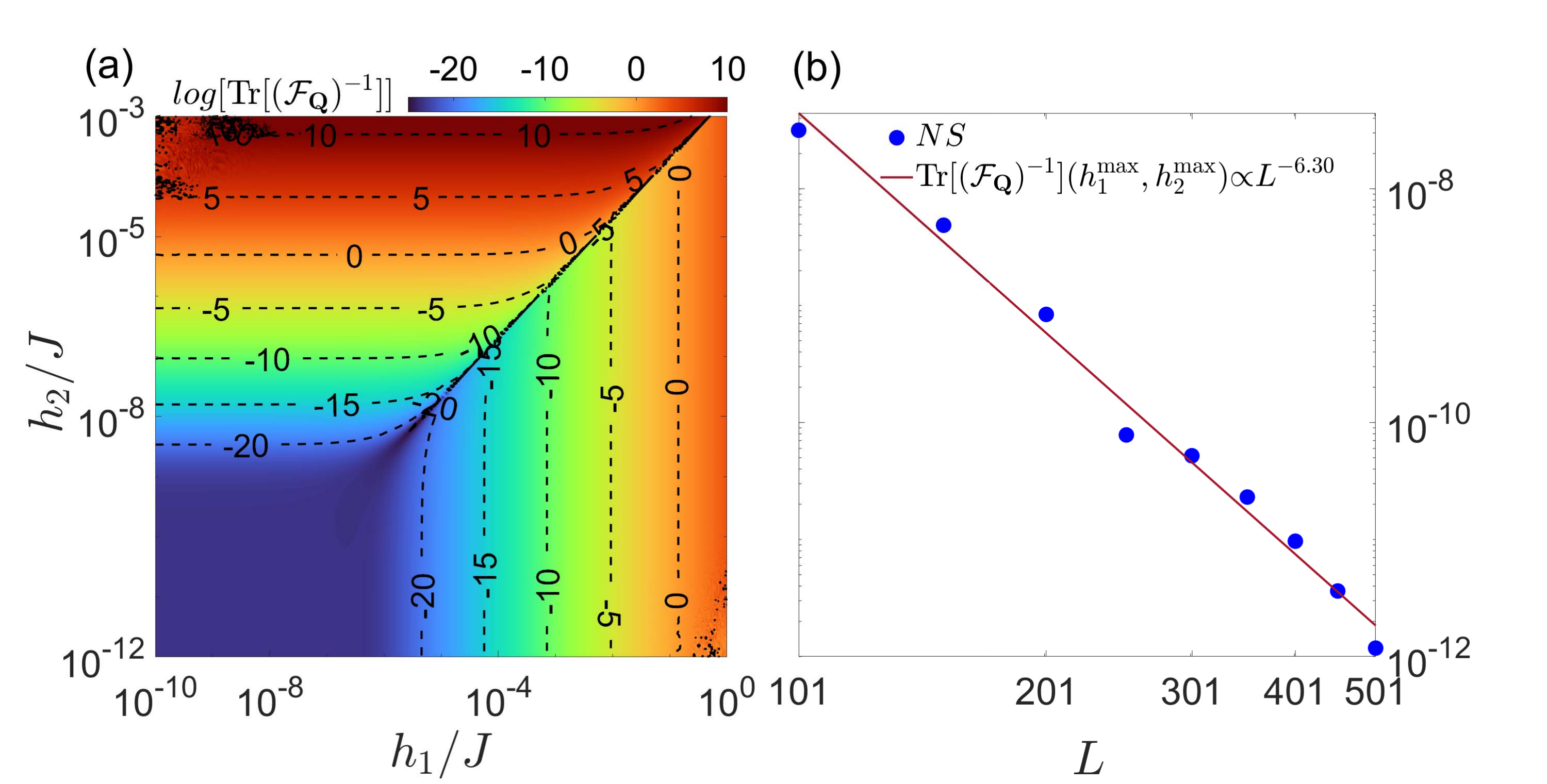} 
\includegraphics[width=\linewidth]{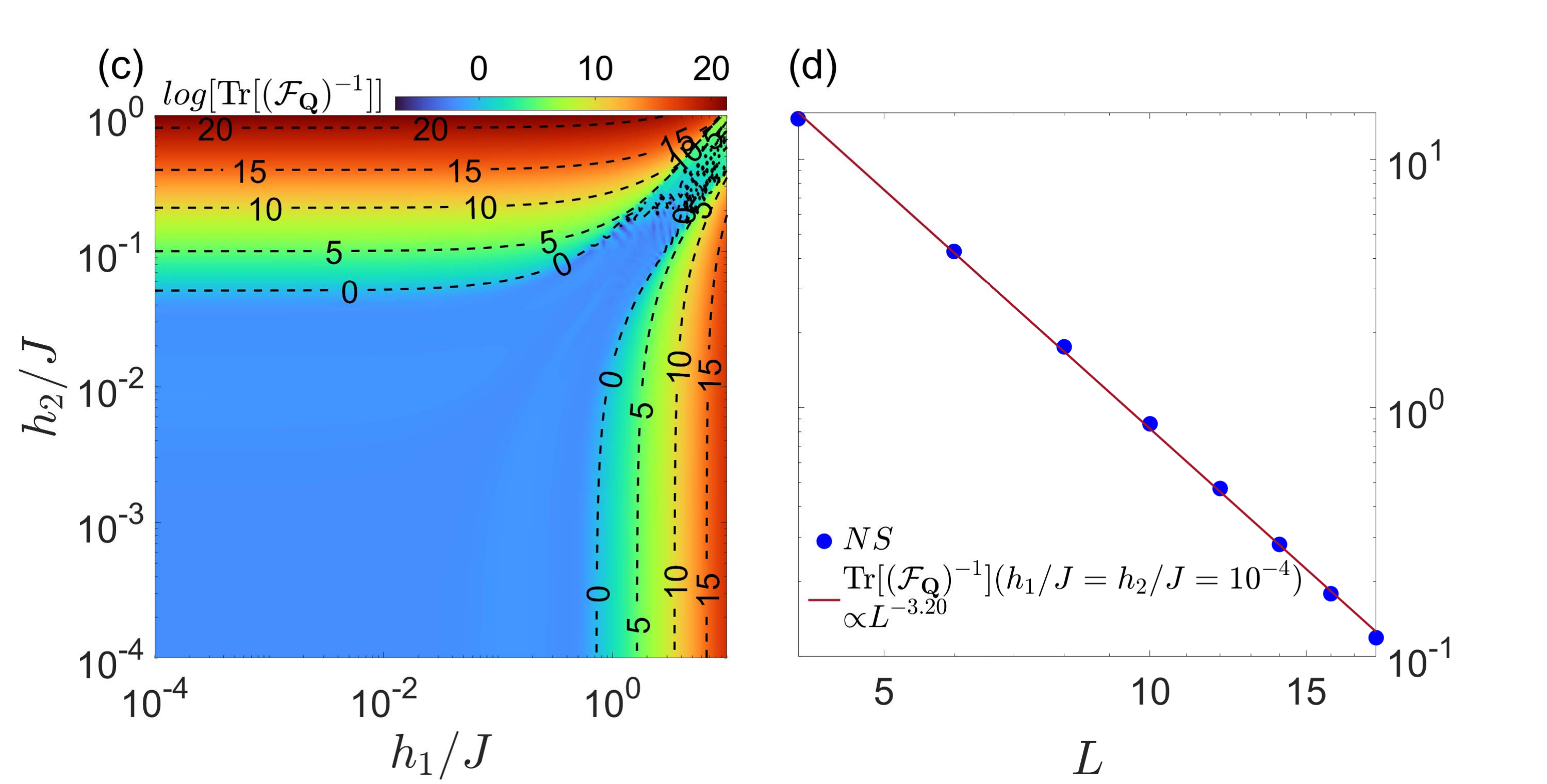} 
\caption{ Total uncertainty ${\rm Tr}[(\boldsymbol{\mathcal{F}_{Q}})^{-1}]$ for the equally weighted multi-parameter estimation, as a function of $h_1$ and $h_2$ for (a) single-particle and (c) many-body interacting probes prepared in the corresponding ground state. 
${\rm Tr}[(\boldsymbol{\mathcal{F}_{Q}})^{-1}]$ versus $L$ for (b) single-particle and (d) many-body interacting probes 
at $(h_1^{\max},h_2^{\max})$. The NS are well described by a fitting function as ${\rm Tr}[(\boldsymbol{\mathcal{F}_{Q}})^{-1}]{\propto}L^{-\beta}$ with $\beta=6.3$ and $\beta=3.2$ for single-particle and many-body interacting probes, respectively.
}\label{fig:SE}
\end{figure}
Up to now, we only focus on the performance of the Stark probes by studying the QFI matrix elements.
Despite the differences in these elements, their behavior in the extended and localized phases, as well as at the transition point look similar.
This hints that the Stark probe can potentially realize a simultaneous estimation scenario.
Back to Eq.~(\ref{Eq.WeightedQRB}), one can establish an equally weighted multi-parameter estimation by choosing $\boldsymbol{W}{=}\mathbb{I}$ and calculating $\text{Tr}[\boldsymbol{\mathcal{F}_{Q}}^{-1}(\boldsymbol{h})]$ as the ultimate estimation precision that lower bound total uncertainty  
\begin{equation} \label{eq:Cramer_Rao_InvQFI}
\delta h_1^2{+}\delta h_2^2 \geq \text{Tr}[\boldsymbol{\mathcal{F}_{Q}}^{-1}(\boldsymbol{h})] = \frac{(\boldsymbol{\mathcal{F}_{Q}})_{11}+(\boldsymbol{\mathcal{F}_{Q}})_{22} }{(\boldsymbol{\mathcal{F}_{Q}})_{11}(\boldsymbol{\mathcal{F}_{Q}})_{22}-(\boldsymbol{\mathcal{F}_{Q}})_{12}(\boldsymbol{\mathcal{F}_{Q}})_{21}}
\end{equation}
Our results for $\text{Tr}[\boldsymbol{\mathcal{F}_{Q}}^{-1}(h_1,h_2)]$ is presented in Fig.~\ref{fig:SE}(a) for single-particle probe of size $L{=}501$.
Obviously, the lowest uncertainty is obtained in the extended phase as well as along the symmetric line.
In Fig.~\ref{fig:SE}(b), we report the lowest values of $\text{Tr}[\boldsymbol{\mathcal{F}_{Q}}^{-1}]$ which happens at   $(h_{1}^{\max},h_{2}^{\max})$ for system of different sizes. 
The behavior of the numerical results (markers) can approximately be described by the fitting function $\text{Tr}[\boldsymbol{\mathcal{F}_{Q}}^{-1}]{\propto}L^{-\beta}$ with $\beta{=}6.30$.
\\

Regarding many-body probes, in Fig.~\ref{fig:SE}(c) we plot 
$\text{Tr}[\boldsymbol{\mathcal{F}_{Q}}^{-1}]$ for a many-body probe of size $L{=}18$.
Clearly, the lowest uncertainty can be obtained in the delocalized regime. In Fig.~\ref{fig:SE}(d) we report $\text{Tr}[\boldsymbol{\mathcal{F}_{Q}}^{-1}]$ (markers) of different probe sizes obtained for parameters deep in the delocalized phase, namely $h_1{=}h_2{=}10^{-4}J$. 
The solid line is the best fitting function of the form $\text{Tr}[\boldsymbol{\mathcal{F}_{Q}}^{-1}]{\propto}L^{-\beta}$ with $\beta{=}3.20$. 

%%%%%%%%%%%%%%%%%%%%%%%%%%%%%%%%%%%%%%%%%%%%%%%%%
\section{Conclusion}\label{ConClusion}
The capability of Stark probes has already been identified for measuring linear tiny gradient fields with unprecedented precision, well beyond the capacity of most known critical-based sensors. In this work, we investigate the 
ability of the ground state of these probes in sensing nonlinear Stark fields. For nonlinear form of the field, i.e.  $V_{i}{=}h i^\gamma$, the probe performs with super-Heisenberg scaling precision, in the extended (nonlocalized) phase. This is  quantified by the scaling of the QFI as $L^\beta$. 
We analytically find that the relationship between the scaling exponent $\beta$ and the nonlinearity parameter $\gamma$ follows a universal behavior $\beta{=}a\gamma{+}b$ with $a{>}0$. The results are backed by numerical simulations. In addition, the critical exponents of the transition between the extended and the localized phases are also modified as ${\simeq}a\gamma{+}b$, although the coefficients $a$ and $b$ vary for each exponent.

We also investigate the two parameter estimation problem in the Stark systems when a linear field $h_1$ and a quadratic field $h_2$ affect the system simultaneously.   
We depict the phase diagram of the system based on the QFI matrix elements to illustrate the phase transition driven by $(h_1,h_2)$.  
The elements of the QFI matrix scale algebraically as $L^\beta$ where the scaling exponent $\beta$ shows the same dependence on nonlinearity as for the case of single parameter probes, namely  $\beta{=}a\gamma{+}b$ with $(a,b){\simeq}(2,4)$.
Therefore, one obtains $\beta{\simeq}6$ and $\beta{\simeq}8$ for sensing $h_1$ and $h_2$, respectively. 
In many-body interacting probes, the corresponding scaling exponent for estimating $h_1$ and $h_2$ are obtained as $\beta{\simeq}3$ and $\beta{\simeq}5$, respectively. 
We also demonstrate that  
the CFI, obtained by a set of simple and experimentally available measurements, can closely saturate the QFI matrix. Finlay, we show that quantum-enhanced sensitivity remains valid even if one incorporates the preparation time in our figure of merit. 

\section{Acknowledgement}
 A.B. acknowledges support from the National Natural
Science Foundation of China (Grants No. 12050410253,
No. 92065115, and No. 12274059), and the Ministry of Science and Technology of China (Grant No. QNJ2021167001L). R. Y. acknowledges support from the National Science Foundation of China for the International Young Scientists Fund (Grant No. 12250410242).
A. C. acknowledges financial support from European Union NextGenerationEU through project PRJ-1328 “Topological atom-photon interactions for quantum technologies" (MUR D.M. 737/2021) and project PRIN 2022-PNRR P202253RLY 'Harnessing topological phases for quantum technologies'.

\end{document}